\documentclass[]{interact}
\usepackage{hyperref}
\usepackage{epstopdf}
\usepackage{subfigure}

\usepackage{natbib}
\bibpunct[, ]{(}{)}{;}{a}{}{,}
\usepackage{xcolor}
\renewcommand\bibfont{\fontsize{10}{12}\selectfont}

\theoremstyle{plain}
\newtheorem{theorem}{Theorem}[section]

\theoremstyle{definition}
\newtheorem{definition}[theorem]{Definition}

\theoremstyle{remark}
\newtheorem{remark}{Remark}

\begin{document}

\articletype{ARTICLE TEMPLATE}

\title{Bayesian Hierarchical Modeling for Predicting Spatially Correlated Curves in Irregular Domains: A Case Study on PM$_{10}$ Pollution}

\author{
\name{Alvaro Alexander Burbano Moreno \textsuperscript{a}\thanks{CONTACT A.~N. Author. Email: aamoreno@unicamp.br} and Ronaldo Dias\textsuperscript{b}}
\affil{\textsuperscript{a} \textsuperscript{b} Instituto de Matemáticas, Estatística e Computação Científica. Universidade Estadual de Campinas, Campinas, SP, Brazil.}
}

\maketitle

\begin{abstract}
This study presents a Bayesian hierarchical model for analyzing spatially correlated functional data and handling irregularly spaced observations. The model uses Bernstein polynomial (BP) bases combined with autoregressive random effects, allowing for nuanced modeling of spatial correlations between sites and dependencies of observations within curves. Moreover, the proposed procedure introduces a distinct structure for the random effect component compared to previous works. Simulation studies conducted under various challenging scenarios verify the model's robustness, demonstrating its capacity to accurately recover spatially dependent curves and predict observations at unmonitored locations. The model's performance is further supported by its application to real-world data, specifically PM$_{10}$ particulate matter measurements from a monitoring network in Mexico City. This application is of practical importance, as particles can penetrate the respiratory system and aggravate various health conditions. The model effectively predicts concentrations at unmonitored sites, with uncertainty estimates that reflect spatial variability across the domain. This new methodology provides a flexible framework for the FDA in spatial contexts and addresses challenges in analyzing irregular domains with potential applications in environmental monitoring. 
\end{abstract}

\begin{keywords}
Bayesian Hierarchical Modeling; Spatial Functional Data; Irregular Sampling; Spatial Dependence; Bernstein Polynomials.
\end{keywords}

\section{Introduction}
Functional Data Analysis (FDA) has gained considerable attention as a statistical framework for analyzing continuous data variations over a specified interval. Unlike conventional statistical methods that often treat observations as scalars or vectors, which may inadequately capture the complexity of real-world datasets, FDA views data as functions, providing a more comprehensive understanding of their behavior, dynamics, and variability. This approach is suitable for analyzing data collected over time, spatial dimensions, or any continuous domain. Such data examples are found across diverse fields such as economics \citep{muller:2011}, medicine \citep{li:2017,shi:2022}, biology \citep{fu:2019}, and engineering \citep{kim:2021}. Interested readers can refer to introductory overviews of FDA in \cite{ramsay:2002,ramsay:2005}, \cite{ferraty:2006}, and \cite{kokoszka:2017}.

In spatial data analysis, a fundamental concept is neighborhood dependency, recognizing that observations within a spatial context are often interrelated, with values at neighboring locations exhibiting higher similarity than those farther apart \citep{cressie:1993, banerjee:2014}. Incorporating spatial dependency into statistical models is crucial for inference and prediction, particularly in fields such as environmental science, geostatistics, and epidemiology \citep{diggle:2007,lawson:2021}.

The combination of FDA with spatial dependency modeling has resulted in an innovative approach known as Spatial Functional Data (SFD) analysis. Unlike traditional FDA techniques, which treat observations only as functions, SFD incorporates spatial location as an additional dimension. This integration enables researchers to explore dependence and heterogeneity in functional data more effectively, offering insights into underlying processes \citep{Jorge:2017, Israel:2020} and \cite{mateu:2021}.

Several recent studies have introduced various statistical methodologies for analyzing multivariate spatial data \citep{brown:1994,ver:1998,gelfand:2005,datta:2016}. However, research on the FDA in cases where functions exhibit spatial dependence is limited. Within the frequentist approach to statistical analysis, notable works have concentrated on SFD. For instance, \cite{Nerini:2010}  suggested a spatial functional linear model for analyzing oceanographic data. \cite{Zhou:2010} introduced mixed effects models for hierarchical functional data with spatial correlation, While \cite{giraldo:2012} proposed a methodology for clustering functional data that are spatially correlated, \cite{jiang:2012} and \cite{romano:2017} offer additional insights. \cite{staicu:2010} proposed a methodology tailored for functional models with a hierarchical structure, where functions at the lowest hierarchy level exhibit spatial correlation. Additionally, \cite{guo:2022} explore clustering methodologies for spatial functional data utilizing functional properties.

One reference to SFD with a Bayesian perspective is \cite{baladandayuthapani:2008}, which employed a Bayesian semi-parametric method using regression splines to handle general between-curve covariance structures. Another study by \cite{zhang:2016} introduced a functional conditional autoregressive (CAR) model for spatially correlated data. Furthermore, \cite{song:2019} proposed novel models based on wavelets for spatially correlated functional data. These models enable the regularization of curves observed over space and the prediction of curves at unobserved sites. \cite{rekabdarkolaee:2019} introduced a novel multivariate space-time functional model with spatially varying coefficients.\cite{white:2022} present a non-stationary model and address challenges and limitations in functional regression within the spatial functional data framework. Finally, \cite{korte:2022} delves into multivariate functional data analysis applied to space-time data.

The SFD employs nonparametric methods for smoothing functional data, with both classical and Bayesian approaches adopting this strategy \citep{hollander:2013}. The term "nonparametric" indicates the absence of assumed specific forms for the underlying functions describing the data, offering enhanced flexibility for capturing intricate patterns. Several prevalent techniques utilized for this purpose in research include kernels \citep{wand:1994}, splines \citep{boor:2001}, and wavelets \citep{vidakovic:2009}.

In the FDA, researchers have developed various methodologies to tackle the issue of irregular sample spacing. Notable authors such as \cite{james:2000}, \cite{morris:2006}, \cite{thompson:2008}, \cite{Gareth:2010}, along with \cite{liu:2012}, have investigated the application of mixed models. These models integrate functional principal components and basis function techniques such as B-splines and Wavelets, among others, to effectively address this challenge. In the SFD literature, most work focuses on equally spaced measurements. However, recent contributions, such as that of \cite{Philip:2021}, introduced a spatial functional approach to estimate monotonic snow density curves using non-uniformly spaced data. Additionally, \cite{Burbano-Moreno:2024a, Burbano-Moreno:2024b} presents a novel Bayesian Gaussian functional model for analyzing spatially correlated curves. This model employs B-splines smoothing techniques and Bernstein Polynomials (BP) and features a random effect component with an autoregressive structure to account for possible associations between serial data caused by irregular spacing.


The main objective of this paper is to generalize and extend the Bayesian Functional Model proposed in \cite{Burbano-Moreno:2024a, Burbano-Moreno:2024b}, designed to treat curves with spatial dependence. In this context, we are interested in a set of $m$ curves, each composed of discrete points $t_{1j}, t_{2j}, ...,t_{n_{i}j}$ that are ordered and irregularly spaced in a continuous domain. It is important to note that the number of observations and the irregular spacing scheme can vary between curves, which previous studies have not addressed. Another of the key differences between this work and that of \cite{Burbano-Moreno:2024a, Burbano-Moreno:2024b} involves the consideration of a different structure for the random effect component. Unlike their approach, which scales the distances between discrete points in the functional domain to the interval [0,1] to model the dependence between $t_i$ by a standard normal cumulative function, we choose to generalize their approach by incorporating a first-order autoregressive process AR(1) to account for potential associations in the serial data caused by the irregular spacing patterns mentioned earlier. The proposed modeling technique uses the smoothing method with BP basis functions \citep{lorentz:1986,farouki:1987,farouki:1988}. These functions depend only on the degree and the defined interval. The model's unknown parameters are estimated using Bayesian inference, a widely adopted approach that incorporates prior knowledge and deals with uncertainties associated with unknown variables. This approach enables the modeling of dependency patterns and helps draw explicit conclusions during the analysis.

The proposed methodology is suitable for modeling spatial dependencies in irregular domains and filling critical gaps in the analysis of spatial functional data, the study of particulate matter is particularly relevant. Pollution from inhalable particles, such as particulate matter with a diameter of less than 10 micrometers (PM$_{10}$), has received significant attention due to its adverse impacts on public health. These fine particles, capable of penetrating deep into the respiratory system, are associated with various health conditions, including respiratory and cardiovascular diseases and cancer. Monitoring and modeling PM$_{10}$ concentrations in dense urban environments is essential to mitigate these impacts, particularly in cities like Mexico City, where industrial activities and vehicular traffic generate considerable spatial and temporal variability in pollution levels.

This paper is structured as follows: Section \ref{sec:2} introduces the SFD model, which incorporates the spatial dependence structure, including the association motivated by irregular spacing and the BP smoothing technique. Section \ref{Cap:3} shows a simulation study based on artificial data and a Monte Carlo (MC) scheme to explore the performance of the proposed model in terms of inference and prediction. Section \ref{Cap:4} presents a practical application for analyzing functional data collected from PM$_{10}$ pollution particle monitoring stations in Mexico City. Section \ref{Cap:5} outlines the main conclusions.
\section{Theoretic Framework}\label{sec:2}
\subsection{Bernstein Polynomial Basis}\label{sec:2.1}

Polynomials are an attractive class of functions for various scientific and engineering computations. They are concisely represented by coefficients on a suitable basis and are amenable to efficient evaluation by simple algorithms. The set of polynomials is closed under the arithmetic operations of addition, subtraction, multiplication, differentiation, integration, and composition. The approximative capabilities of polynomials are also of great practical interest in applications. Perhaps the most fundamental result in this context is the theorem of Weierstrass, which is stated below \citep{davis:1975}:
    \begin{theorem}
    Let $g$ be a real and continuous function defined on a compact interval $[a,b]$. Then for each $\varepsilon>0$ there exists a polynomial $p$ (which depends on $\varepsilon$) such that
    \begin{align*}
        |g(t)-p(t)|<\varepsilon\hspace{0.4cm}\text{for each $t$ of $[a,b]$.}
    \end{align*}
    \end{theorem}
    In other words, it is possible to uniformly approximate any continuous function $g$, defined on a polynomial's closed interval $[a,b]$. An elegant constructive proof of this theorem was published in 1912, in which Bernstein's polynomial basis was first introduced; for more details, see \cite{bernvstein:1912} and \cite{lorentz:1986}.
    
   \begin{definition}[Bernstein basis functions] Let $p$ denote any non-negative integer, and suppose $[a,b]$ is a bounded interval in $\mathbb{R}$. The polynomials 
    \begin{align}\label{ec:2.1}
	b_{\,r}^{\,p}(t)=\binom{p}{r}\left ( \frac{t-a}{b-a} \right )^{r}\left (1-\frac{t-a}{b-a} \right )^{p-r}, \quad\text{for}\quad r=0,\ldots,p,
	\end{align}
	are called the Bernstein polynomials of degree $p$ $(order\, p+1)$ with respect to the interval $[a,b]$.
    \end{definition}
    
    \begin{remark} 
    The domain of the Bernstein basis polynomials can be defined on the interval $[0,1]$ without loss of generality, replacing
    \begin{align}\label{ec:2.2}
    x=\frac{t-a}{b-a},\quad a\leq t\leq b,
    \end{align}
    or equivalently,
    \begin{align}\label{ec:2.8}
    t=(b-a)x+a,\quad 0\leq x\leq 1.
    \end{align}
    By using \eqref{ec:2.2}, and \eqref{ec:2.8}, it is observed from \eqref{ec:2.1} that
    \begin{align}\label{ec:2.3}
	b_{\,r}^{\,p}(t)=\binom{p}{r} x^{r}\left (1-x \right )^{p-r}, \quad\text{for}\quad r=0,\ldots,p.
	\end{align}
 \end{remark}
 \begin{remark}
For any non-negative integer $p$, and bounded interval $[a,b]\subset\mathbb{R}$, the corresponding Bernstein polynomials, as defined by \eqref{ec:2.1}, satisfy:
\begin{itemize}
    \item[i.] Recursive generation. The basis of degree $p$ may be generated from the basis of degree $p-1$
    \begin{align*}
        b_{\,r}^{\,p}(t)=&\,\binom{p}{r}\left ( \frac{t-a}{b-a} \right )^{r}\left (1-\frac{t-a}{b-a} \right )^{p-r}\\[0.15cm] \nonumber
        =&\,\left (1-\frac{t-a}{b-a} \right )b_{\,r}^{\, p-1}(t)+\left (  \frac{t-a}{b-a}\right )b_{\,r-1}^{\,p-1}(t).
    \end{align*}
    \item[ii.] \begin{align*}
	b_{\,r}^{\,p}(a)= \begin{cases}
	1	& \text{ if }\; r=0, \\ 
	0	&\; \text{otherwise}, 
	\end{cases}\qquad \text{and}\qquad
 b_{\,r}^{\,p}(b)= \begin{cases}
	1	& \text{ if }\; r=p, \\ 
	0	&\; \text{otherwise}. 
	\end{cases}
\end{align*}
    \item[iii.] The positivity and partition of unity properties on $[a,b]$
\begin{align*}
        b_{\,r}^{\,p}(t)\geq 0,\quad r=0,\ldots,p\quad \text{and}\quad \sum_{r=0}^{p}b_{\,r}^{\,p}(t)=1.    
\end{align*}
        \item[iv.] Let $\pi_{p}$ be a finite-dimensional linear space, such that $dim(\pi_{p})=p+1$. Then, the polynomial sequence $\left \{ b_{\,r}^{\,p}(t),\; r=0,\ldots,p \right \}$ is a basis for $\pi_{p}$. 
\end{itemize}
\end{remark}
 
\begin{figure}[!h]	
	\centering
	\includegraphics[width=1\textwidth]{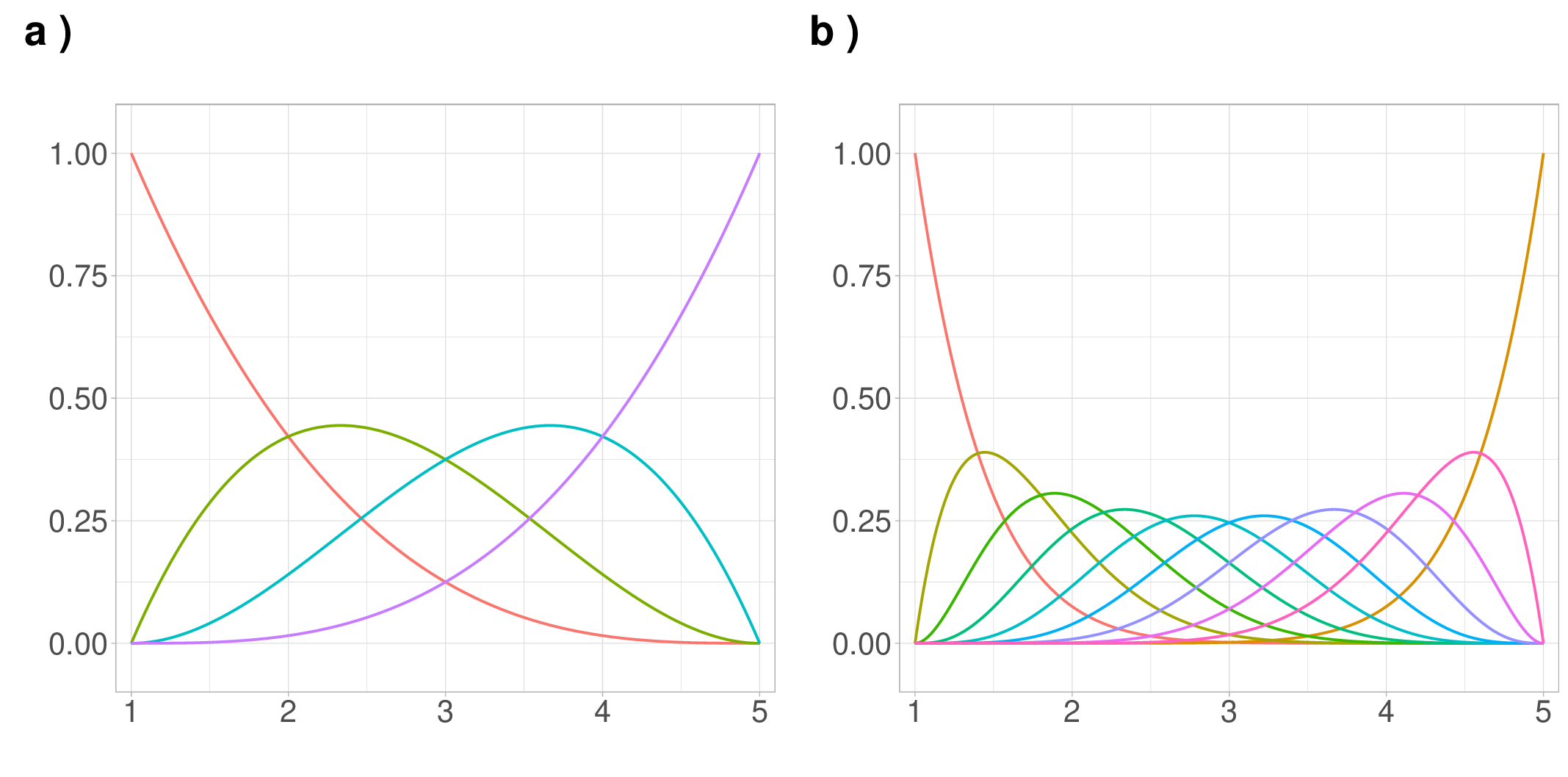}
	\caption{Illustration of Bernstein basis for $p=3$ (Panel a) and $p=9$ (Panel b).}
	\label{fig:2.1}
\end{figure}

An illustration of the Bernstein basis can be seen in Figure \ref{fig:2.1}. The vector of basis $\mathbf{b}^{p}(t)=\left ( b_{\,0}^{\,p}(t),\ldots, b_{\,p}^{\,p}(t) \right )$ has a weight role, which varies with $t$, as shown in both panels of Figure \ref{fig:2.1}. Thus, the approximation of the target function $g(\cdot)$ is weighted by $p+1$ values coming from the basis vector. It is clear that when $p= 9$, more information is available to weigh the function values, resulting in a more accurate approximation.
\subsection{Statistical Model}\label{sec:2.2}
A spatial functional process is described as $\left\lbrace X_{s}:s\in\mathcal{D}\subset\mathbb{R}^{d}\right\rbrace $, where $X_{s}$ are the  functional random variables, located at the site $s$ in the $d$-dimensional Euclidean space, usually $d=2$ or $3$. Each $X_{s}$ is defined on the interval $T=[a,b]\subseteq \mathbb{R}$ and is assumed to belong to a Hilbert space of square-integrable functions, that is, in 
$$L^{2}(T)=\left\{X_{s}: T\longrightarrow \mathbb{R},\textnormal{ such that} \int_{T}X_{s}(t)^{2}dt<\infty  \right\},$$
with the inner product $\left\langle X_{s}, X_{s^{\ast }}  \right\rangle=\int_{T}X_{s}(t)X_{s^{\ast }}(t)dt $.

Let $\left\lbrace X_{s_{1}}(t),\ldots,X_{s_{m}}(t)\right\rbrace$ be a sample of $m$ non-independent curves that are indexed by a domain $t\in T$ at each spatial location $s_{j}$, $j=1,\ldots,m$. The functions are observed through a finite set of discrete points, denoted as $ t_{ij},\, i=1,\ldots,n_{j}$. It's essential to note that these measurements are frequently affected by noise. Considering a specific location represented by $s_{j}$, we assume that the observed functions are generated according to the following model
\begin{equation}
\label{ec:3.1}
	Y_{s_{j}}(t_{ij})=X_{s_{j}}(t_{ij})+\delta_{ij} +\epsilon_{s_{j}}(t_{ij}),
\end{equation}
where $\delta_{ij}$ represents a random effect, and $\epsilon_{s_{j}}(t_{ij})$ are independent random errors associated with each fixed point $t_{ij}$. These errors have a mean of zero, i.e., $E(\epsilon_{s_{j}}(t_{ij}))=0$, and a common unknown variance, denoted as $Var(\epsilon_{s_{j}}(t_{ij}))=\tau^{2}$. Henceforth, we assume that each realization $X_{s_{j}}(t_{ij})$ of an underlying random function $X_{s}(t)$ can be adequately represented by a finite set of BP basis functions denoted as $b_{\,0}^{\,p}(t_{ij}),\ldots,b_{\,p}^{\,p}(t_{ij})$ to obtain a reasonable approximation. Consequently, each curve can be expanded into this basis as follows  
\begin{equation}\label{ec:3.2}
    X_{s_{j}}(t_{ij})\approx\sum_{r=0}^{p}\theta_{rj}b_{\,r}^{\,p}(t_{ij}),\hspace{0.3cm} r=0,\ldots,p, 
\end{equation}
for $p\in \mathbb{N}$. The basis coefficients, denoted as $\theta_{r,j}$, are crucial in constructing the correlation between curves. Drawing inspiration from the works of \cite{liu2017} and \cite{aguilera2017}, it is assumed that $\theta_{r,j}$ is associated across different locations $j$ for each $r$, but not across different basis function indices $r$. To fully define the spatial correlation structure among $\theta_{r,j}$'s, we introduce the possibility of a non-zero covariance. Specifically, we express this covariance as $Cov(\theta_{r,j},\theta_{r,j^{*}})=C(s_{j},s_{j^{*}})=C(h)$, where $j\neq j^{*}$. Notably, the covariance function is contingent on the locations $s_{j}$ and $s_{j^{}}$ solely through the Euclidean distance $h=\parallel s_{j}-s_{j^{*}} \parallel\in \mathbb{R}$. Taking Equation \eqref{ec:3.2} into account, we can express the model described in Equation \eqref{ec:3.1} as follows
\begin{equation}\label{ec:3.3}
	Y_{s_{j}}(t_{ij})=\sum_{r=0}^{p}\theta_{rj}b_{\,r}^{\,p}(t_{ij})+\delta_{ij} +\epsilon_{s_{j}}(t_{ij}).
\end{equation}
In the context of SFD, a common practice involves linking discrete realizations of each function with finite points $t$ that are regularly spaced along the functional domain. However, a fundamental and distinct aspect of the methodology proposed in this paper is the use of non-equidistant distances between measurement points. In other words, given a sequence of increasing consecutive points $t_{ij}$, denoted by
$d_{ij}=(t_{ij}-t_{(i-1)j})\neq(t_{i^{*}j}-t_{(i^{*}-1)j}) = d_{i^{*}j}$ for some $i \neq i^* \in \left \{2,\ldots,n_{j} \right \}$ are not necessarily equal. In equation \eqref{ec:3.3}, we use the locations' geographic coordinates $s_{j}$ to introduce spatial dependence between the observed functional trajectories. This approach takes into account both the spatial relationships and an autoregressive random effect $\delta_{ij}$, which employs the distances $d_{ij}$ to measure the similarity between discrete measurements closely located within each function $Y_{s_{j}}(t)$. We include this structure to account for the non-equidistant spacing of observations throughout the functional domain. The mathematical structure of $\delta_{j}=\left(\delta_{1j},\ldots,\delta_{n_{i}j}\right)^\top$ is defined below.
\begin{equation}\label{ec:3.4}
    \delta_{ij}=\phi_{ij}\delta_{(i-1)j}+\epsilon_{\delta_{ij}}\hspace{0.3cm}\text{with}\hspace{0.3cm} \phi_{ij}=\exp{(-\eta d_{ij})}. 
\end{equation}
We introduce a sequence of uncorrelated and identically distributed Gaussian random variables in the given specification, denoted by $\epsilon_{\delta_{ij}}$. These variables have a mean of zero and a variance of $\nu^{2}$, for $i=1,\ldots, n_{j}$ and $j=1,\ldots, m$. Regarding $\phi_{ij}$, it has the structure of an exponential correlation function, where $d_{ij}$ represents the interval between two observations and $\eta$ is the parameter that controls the correlation's decay rate. A higher value of $\eta$ results in a faster decay, implying a lower correlation between farther-apart observations. Note that an autoregressive structure is defined to associate the random effects in $\delta$. In other words, it ensures that the level of relationship between $\delta_{ij}$ and $\delta_{(i-1)j}$, concerning the positions $t_{ij}$ and $t_{(i-1)j}$ of the functional domain, is controlled by the coefficient $\phi$. Assume $d_{0j} = 0$ (so $\phi_{1j}= 1$) and $\delta_{0j} = 0$; it implies that $\delta_{1j}\sim N(0, \nu^{2})$.
\subsection{Bayesian Hierarchical Models} \label{sec:2.3}
The choice of a Bayesian hierarchical model is crucial when dealing with complex data sets that exhibit both within-group and between-group variation. By structuring the model hierarchically, we can capture these different sources of variability and estimate parameters at each level, providing more robust and interpretable results.

Let's consider the $i$th discrete observation for the $j$th curve, denoted as $Y_{s_{j}}(t_{ij})$, along with a set of BP basis functions $\left \{b_{\,0}^{\,p}(t_{ij}),\ldots, b_{\,p}^{\,p}(t_{ij})\right \}$. The model incorporates parameters $\theta_{rj}$, $\delta_{ij}$, and $\tau^{2}$ (see Subsection \ref{sec:2.2}). We can construct the following Bayesian hierarchical model: sampling for $i=1,\ldots,n_{j}$ and $j=1,\ldots,m$,

\begin{align}\label{ec:3.5}
&Y_{s_{j}}(t_{ij})|\,\theta_{rj}, \delta_{ij}, \tau^{2}\sim\, N\left(\sum_{r=0}^{p}\theta_{rj}b_{\,r}^{\,p}(t_{ij})+\delta_{ij},\, \tau^{2} \right);\nonumber\\[0.2cm]
	&\bm{\theta}_{r}|\,\mu_{\theta_{r}},  {\bm \Sigma_{m}}\sim\, N_{m}\left( \mu_{\theta_{r}}{\bm 1_{m}},{\bm \Sigma_{m}}\right),\quad \text{where $\bm{\theta}_{r}=\left(\theta_{r1},\ldots,\theta_{rm} \right)^{\top}$};\nonumber\\[0.15cm]
	&\delta_{ij}|\,\delta_{(i-1)j}, \eta,\nu^{2}\sim\, N\left (\phi_{ij}\,\delta_{(i-1)j},\,\nu^{2} \right ),\quad\text{with $\phi_{ij}=\exp{(-\eta\, d_{ij})}$};\\[0.15cm]
    &\delta_{1j}|\,\nu^{2}\sim\, N\left(0,\nu^{2} \right);\nonumber\\[0.15cm]
	&\mu_{\theta_{r}}\,\sim\, N\left(o,v^2\right);\nonumber \\[0.15cm]
 &\varphi\sim IG\left(a_{\varphi},b_{\varphi} \right);\nonumber\\[0.15cm]
    &\eta\,\sim\, IG\left(a_{\eta}, b_{\eta}\right);\nonumber \\[0.15cm]
    &\kappa^{2}\,\sim\, IG\left(a_{\kappa^2},b_{\kappa^2} \right);\quad\nu^{2}\,\sim\, IG(a_{\nu^2},b_{\nu^2}); \quad\tau^{2}\,\sim\, IG\left(a_{\tau^2},b_{\tau^2} \right)\nonumber.
\end{align}
In this research, we are utilizing the Gaussian covariance function as described by \cite{banerjee:2014}, represented as $C(h)=\kappa^{2}\exp(-(\varphi\,h)^{2})$, to model an isotropic spatial process within the $m$-order covariance matrix, denoted as $\bm{\Sigma}_{m}$. In this context, $\kappa^{2}$ signifies spatial variation, while $\varphi$ is the spatial decay parameter. The symbol $h$ represents the Euclidean distance between two locations $s_{j}$ and $s_{j^{*}}$. It is crucial to emphasize that when the locations are close in $\mathbb{R}^d$, the basis coefficients tend to show similarity, resulting in curves with similar shapes, where $C(h) \approx \kappa^{2}$. Conversely, as the distance between these locations increases, the dissimilarity in the curves also increases, leading to $C(h) \approx 0$. For the spatial parameters $\kappa^{2}$ and $\varphi$, we are using prior Inverse Gamma distributions (IG). This choice is common in Bayesian hierarchical models due to the IG conjugacy property, simplifying subsequent calculations. We use Gaussian priors for the coefficients $\bm{\theta}_{r}$ towards a mean value $\mu_{\theta_{r}}$ based on our prior knowledge or assumptions about their distribution. By representing $\bm{\theta}_{r}$ with a multivariate normal distribution, we can account for potential correlations between the coefficients across different locations, capturing spatial dependencies in the model. In this context, $\bm{\theta}_{r} = \left(\theta_{r1},\ldots,\theta_{rm} \right)^{\top}$ represents a vector containing the values of the $r$-th coefficient at the $m$ locations, and $\mu_{\theta_{r}}$ is the mean of the $r$-th basis coefficient. Additionally, utilizing a Gaussian prior for $\mu_{\theta_{r}}$ introduces regularization to the mean of the coefficients, enabling control over their overall magnitude and preventing overfitting to the data. The hyperparameters $o$ and $v^{2}$ determine the center and spread of the prior distribution, respectively, and can be selected based on prior knowledge. To capture an autoregressive process that depends on the distance $d_{ij}$ between consecutive discrete points in the continuous domain, we use a Normal prior with mean $\phi_{ij}\,\delta_{(i-1)j}$. This formulation reflects that past effects influence future effects. The priors for $\eta$, $\nu^2$, and $\tau^{2}$ are IG, reflecting uncertainty about the parameters controlling the smoothness, the random effect variance, and the observations.

\section{Simulation Studies}\label{Cap:3}
To assess the effectiveness of the methodology introduced in Section \ref{sec:2}, we conducted two simulation studies using synthetic data. These studies aim to evaluate the performance of the BP basis function and the integration of the autoregressive random effect component for managing irregular spacing. Specifically, we analyze how well the proposed approach recovers spatially dependent sets of curves, as detailed in Subsection \ref{sec:3.1}. Additionally, we examine its effectiveness in predicting curves at geographic locations not included in the training data, as discussed in Subsection \ref{sec:3.2}. The model is implemented using the \texttt{Stan} programming language \citep{stan:2023}, which facilitates full Bayesian inference for continuous variable models via Markov Chain Monte Carlo (MCMC) methods, specifically the No-U-Turn sampler (NUTS), which is an adaptive form of Hamiltonian Monte Carlo (HMC) \citep{Hoffman2014}. The implementation of the model employs the \texttt{R} programming language \citep{R:2023}.

\subsection{First Study: Curve Recovery Under Spatial Correlations}\label{sec:3.1}
\begin{figure}[!hbt]	
	\centering
	\includegraphics[width=1\textwidth]{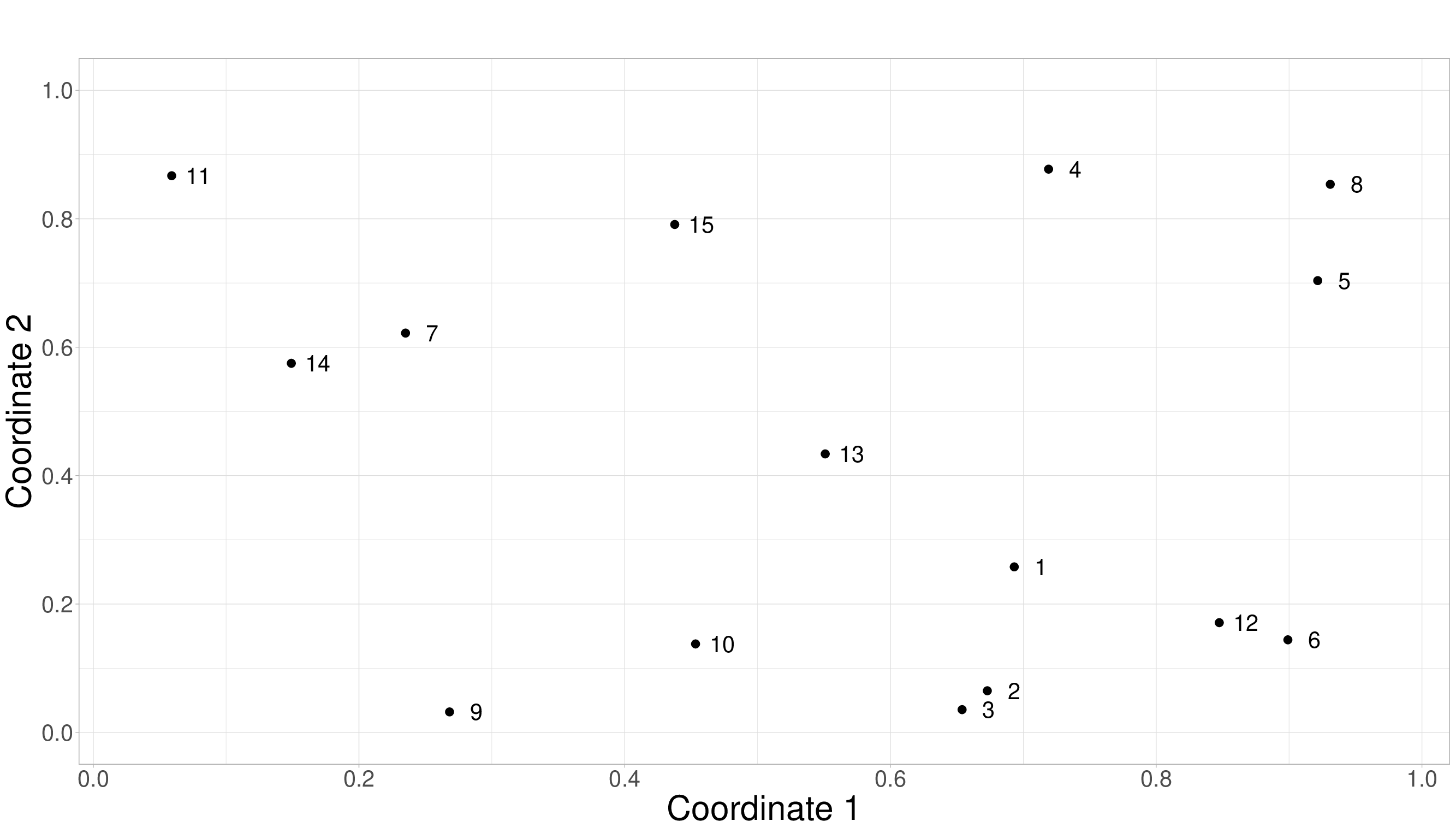}
	\caption{Grid of simulated sites.}
	\label{fig:3.1}
\end{figure}

Consider a scenario with 15 specific geographic locations, depicted in Figure \ref{fig:3.1}. At each location, 200 discrete points are selected along the continuous domain of the function $t$. These points are associated with a measure $Y$, forming ordered pairs $(t, Y)$ that define the corresponding functions. It's important to note that the spacing between the points is irregular and varies for each function, meaning it doesn't follow a consistent pattern. Let us assume, without loss of generality, that the distances between discrete points in the functional domain are within the interval $(0,1)$. We will examine two situations: In the first case, we will use a Uniform distribution $U(0,1)$, and in the second, from a $Beta(1,2)$. The first option will result in an equal number of small and large distances, while the second will produce more small distances than large ones.

\vspace{5pt}
\noindent \textbf{Synthetic Data}
\vspace{5pt}

The generation of the synthetic data $Y_{s_{j}}(t_{ij})$, with $j = 1,\cdots, 15$ and $i=1,\ldots,n_{j}$, follows the model structure proposed in Equation \eqref{ec:3.5}, and is denoted as $\mathcal{M}_{BP_{p};\,\delta}$, which indicates a BP model of degree $p$ (order $p+1$) that includes the random effect $\delta$. The selection of the appropriate number of BP basis functions to use in a given dataset depends on various factors, such as the complexity of the data and the desired modeling accuracy. Although there is no fixed rule to determine the exact number of basis functions, it is generally accepted that increasing the value of $p$ increases flexibility to obtain a reasonable approximation. However, increasing $p$ also increases the complexity of the resulting polynomials, which can lead to numerical problems and higher computational costs. This study utilized a BP with $p=3$, corresponding to four basis functions. It's important to emphasize that the model used to create the data is identical to the one used to fit the datasets.

The coefficients of the basis expansions, denoted by ${\bm{\theta}_{r}}=(\theta_{r1},\cdots, \theta_{r15})^\top$, have been derived from a normal distribution with mean vector $\mu_{\theta_{r}}\cdot{\bm{1_{15}}}$ and covariance matrix ${\bm{\Sigma}_{15}}$, where ${\bm{1_{15}}}$ represents a vector of ones with dimensions $15 \times 1$. The specific mean values are $\mu_{\theta_{0}}=3$, $\mu_{\theta_{1}}=29$, $\mu_{\theta_{2}}=15$, and $\mu_{\theta_{3}}=7$, with $p = 3$ and $r$ ranging from 0 to 3. The spatial correlation within the data is governed by the covariance matrix $\bm{\Sigma}_{15}$, constructed using the Gaussian covariance function $C(h)=\kappa^{2}\exp({-(\varphi\, h)^2})$, where $h$ denotes the Euclidean distance between the spatial locations $s_{j}$ and $s_{j^*}$. To illustrate the scenario of spatially dependent sets of curves, we consider a spatial variation $\kappa^{2}=2$, along with a moderate spatial correlation level $\varphi=1$. In addition, for the $\delta_{ij}$ structure, we use $\eta=0.2$ and variance $\nu^{2}=0.5$. Finally, the variance of the observations is assumed to be $\tau^{2}=1$.

\vspace{5pt}
\noindent \textbf{Prior Specifications}
\vspace{5pt}

Table \ref{tb:4.2} presents the prior along with the arguments assigned to each parameter and unknown quantity in the model. The pair of values (2,1) chosen for the first four Inverse Gamma distributions indicates high uncertainty about $\varphi$, $\eta$, $\tau^2$, and $\kappa^2$. This configuration results in distributions with a mean of 1 but an undefined variance. The fifth Inverse Gamma is defined by $a_{\nu^2} =3$ and $b_{\nu^2} = 2$, which results in a prior distribution with a mean and variance of 1. This specification reflects an informative expectation about the magnitude of $\nu^{2}$; we expect the variability of the random effects $\delta _{ij}|\delta _{(i-1)j}$ to be moderate, i.e., neither too small nor excessively large.  It's essential because if $\nu^{2}$ were very large, the variability between consecutive observations would be too high, reducing the influence of the mean $\phi _{ij} \delta _{(i- 1 )j}$ on $\delta _{ij}$; see Equation \eqref{ec:3.5}. In addition, the hyperparameter $\mu_{\theta_{r}}$, a Normal prior with a mean of 0 and variance of $50^2$, is proposed to represent low certainty about the real values of the coefficient means.
\begin{table}[!htb]
\centering
 \caption{Prior specifications considered in the simulation study.}
	\begin{tabular}{c|c}\hline
	Hyperparameter& Prior Distribution\\\hline
        $\varphi$&$IG(2,1)$\\[0.15cm]
        $\eta$&$IG(2,1)$\\[0.15cm]
        $\tau^{2}$&$IG(2,1)$\\[0.15cm]
	$\kappa^{2}$&$IG(2,1)$\\[0.15cm]
	$\nu^{2}$&$IG(3,2)$\\[0.15cm]
        $\mu_{\theta_{r}}$&$N(0,50^2)$\\
	\hline
	\end{tabular}
   \label{tb:4.2}
\end{table}

To summarize the inference results, we use the posterior means as estimators. We obtain posterior samples using the MCMC algorithm with HMC dynamics implemented in \texttt{Stan}. We select every tenth iteration from 25,000 runs to reduce autocorrelation while discarding the initial 5,000 runs as the burn-in period. Each parameter is estimated using a single chain, and we visually confirm the convergence of all inspected chains during the analysis.

\vspace{5pt}
\noindent \textbf{Analysis}
\vspace{5pt}

The objective of this initial simulation is to evaluate the effectiveness of the proposed model (as outlined in Equation \ref{ec:3.5}) in recovering the dependent curves produced by the model, along with the corresponding configurations described in Subsection \ref{sec:3.1}.
\begin{figure}[!hbt]	
	\centering
	\includegraphics[width=1\textwidth]{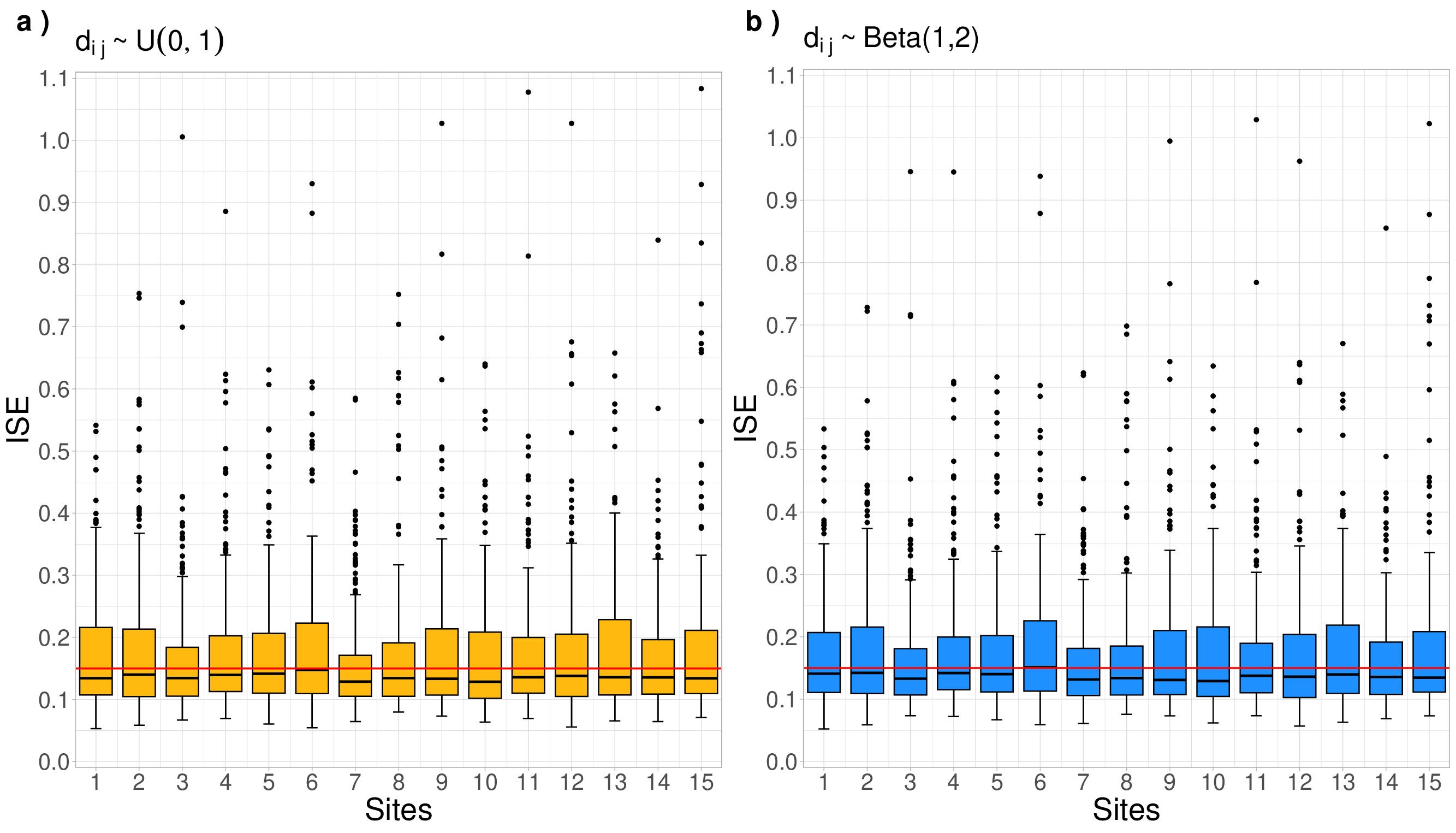}
	\caption{The box plots for the measurements related to 100 ISE. Panel (a) displays observations corresponding to discrete points $t_{ij}$, where the distances follow a uniform distribution $U(0,1)$. In contrast, Panel (b) shows observations at discrete points with distances generated from a $Beta(1,2)$ distribution.}
	\label{fig:3.2}
\end{figure}

In Figure \ref{fig:3.2}, boxplots display the results of 100 measurements of the Integrated Squared Error (ISE) metric (see Appendix \ref{sec:app_metrics}) obtained using the Monte Carlo (MC) scheme. These plots illustrate the approximation performance between smoothed and the target curves across two configurations of irregular domains. In Panel $(a)$, the observations are associated with discrete points $t_{ij}$, where the distances follow a Uniform distribution $U(0,1)$. In this case, there are fewer small distances compared to the $Beta(1,2)$ distribution, indicating a lower level of dependence between the observations of each curve. Conversely, Panel $(b)$ shows observations at discrete points where distances are generated from a $Beta(1,2)$ distribution, resulting in a higher concentration of smaller distances and thus indicating greater dependence between observations within each series. 

The plots suggest that the model $\mathcal{M}_{BP_{3};\,\delta}$ accurately recovers the spatially correlated curves, regardless of the irregularity in the spacing of the domain. Notably, approximately 50\% of the ISE values are below 0.15, as marked by the red line. Additionally, the boxplots in Panel $(a)$ exhibit slightly higher variability than those in Panel $(b)$. In summary, comparing the two panels reveals that the variation in ISE values is more stable under the $Beta(1,2)$ configuration, as indicated by the narrower range of extreme ISE values. On the other hand, the $U(0,1)$ displays a weaker correlation between observations within each series, resulting in a broader range of outliers. This implies that, in some simulations, the approximation performs significantly worse under the uniform distribution.
\subsection{Second Study: Recovery and Prediction of Curves}\label{sec:3.2}
\begin{figure}[!hbt]	
	\centering
	\includegraphics[width=1\textwidth]{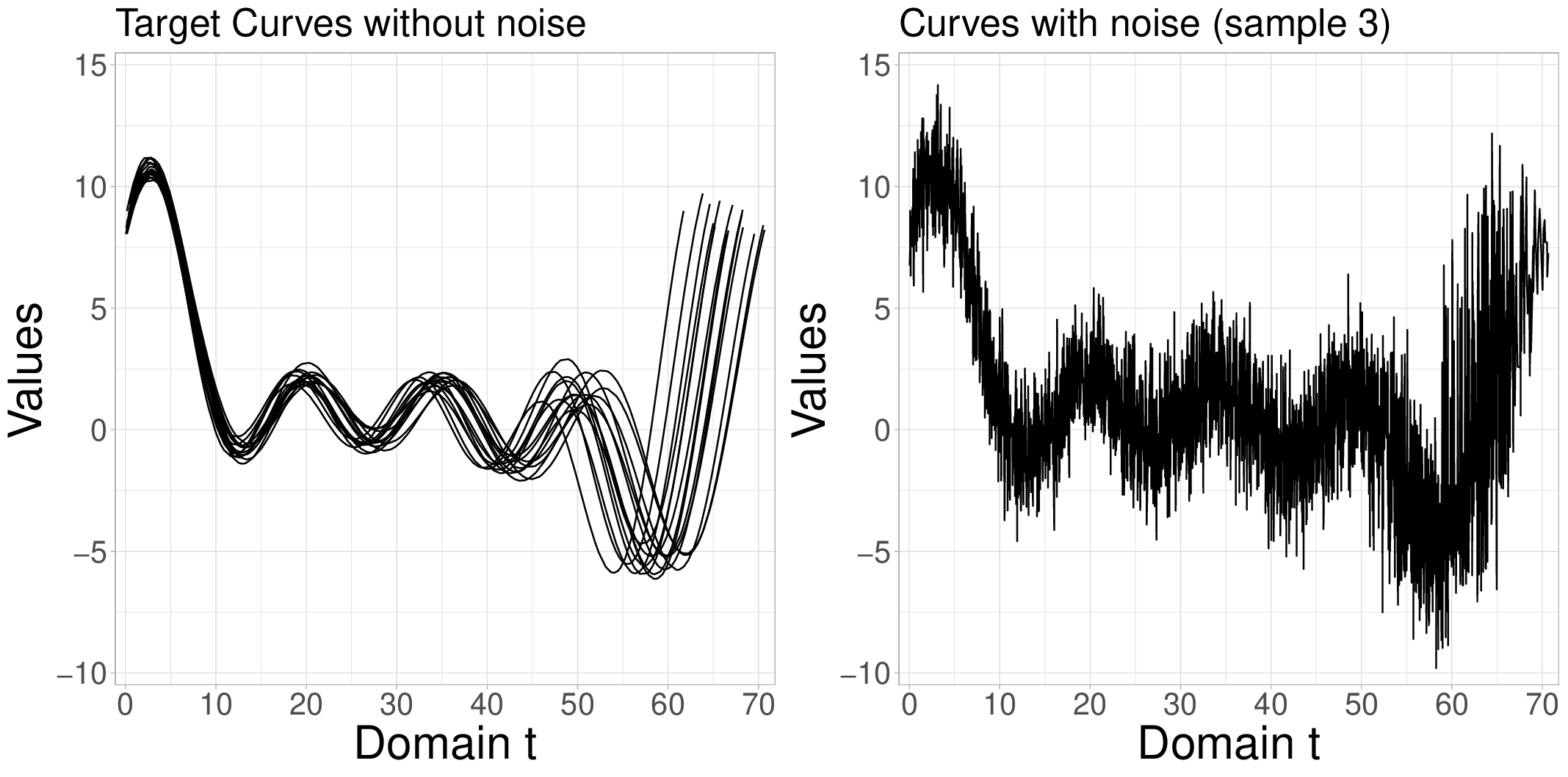}
	\caption{The left panel displays the target curves without noise, while the right panel presents the third sample of the 100 simulated curves in the MC study, where the curves are affected by noise.}
	\label{fig:3.3}
\end{figure}

This study uses the same grid of geographical locations and the functional domain pattern for each curve described in the first simulation study (refer to Figure \ref{fig:3.1}). The dependent curves are generated using a linear combination of 9 Fourier bases.
\begin{align}
    Y_{s_{j}}(t_{ij})=\beta_{0s_{j}}+\sum_{r=1}^{4}\left (\beta_{(2r-1)s_{j}}\sin(r\omega t_{ij})+\beta_{(2r)s_{j}}\cos(r\omega t_{ij}) \right )+\epsilon_{s_{j}}(t_{ij}),
\end{align}
where the constant $\omega$ is related to the period $\mathcal{T}$ by the relation $\omega=2\pi/\mathcal{T}$. We assume tha $\epsilon_{s_{j}}(t_{ij})\sim N(0,\sigma^{2}=2)$ independently for all $i=1,\ldots,200$ and $j=1,\ldots,15$. In addition, $\bm{\beta_{l}}=(\beta_{l\,s_{1}},\ldots,\beta_{l\,s_{15}})^\top\sim N_{15}(10\cdot\bm{1}_{15},\bm{\Sigma_{15}})$, where $l=0,1,\ldots,8$ and $\bm{\Sigma}_{15}$ is defined according to the Exponential covariance function $C(h)=2\exp(-(1\cdot h))$. This study also explores the MC scheme with 100 replicas generated under these conditions. The simulated data sets, with and
without noise, can be seen in Figure \ref{fig:3.3}.

Each synthetic dataset is fitted using the proposed $\mathcal{M}_{BP_{3};\,\delta}$ model and the model without the random effect component $\mathcal{M}_{BP_{3}\,;\,\bullet}$. Both models employ four bases of BP functions of degree 3. The objective is to compare their fitting performances and highlight the importance of considering both spatial dependence and the associations between the discretely observed measurements of each curve. This consideration is motivated by the irregular spacing of the points in the functional domain.

\vspace{5pt}
\noindent \textbf{Analysis}
\vspace{5pt}

\begin{figure}[!hbt]	
	\centering
	\includegraphics[width=1\textwidth]{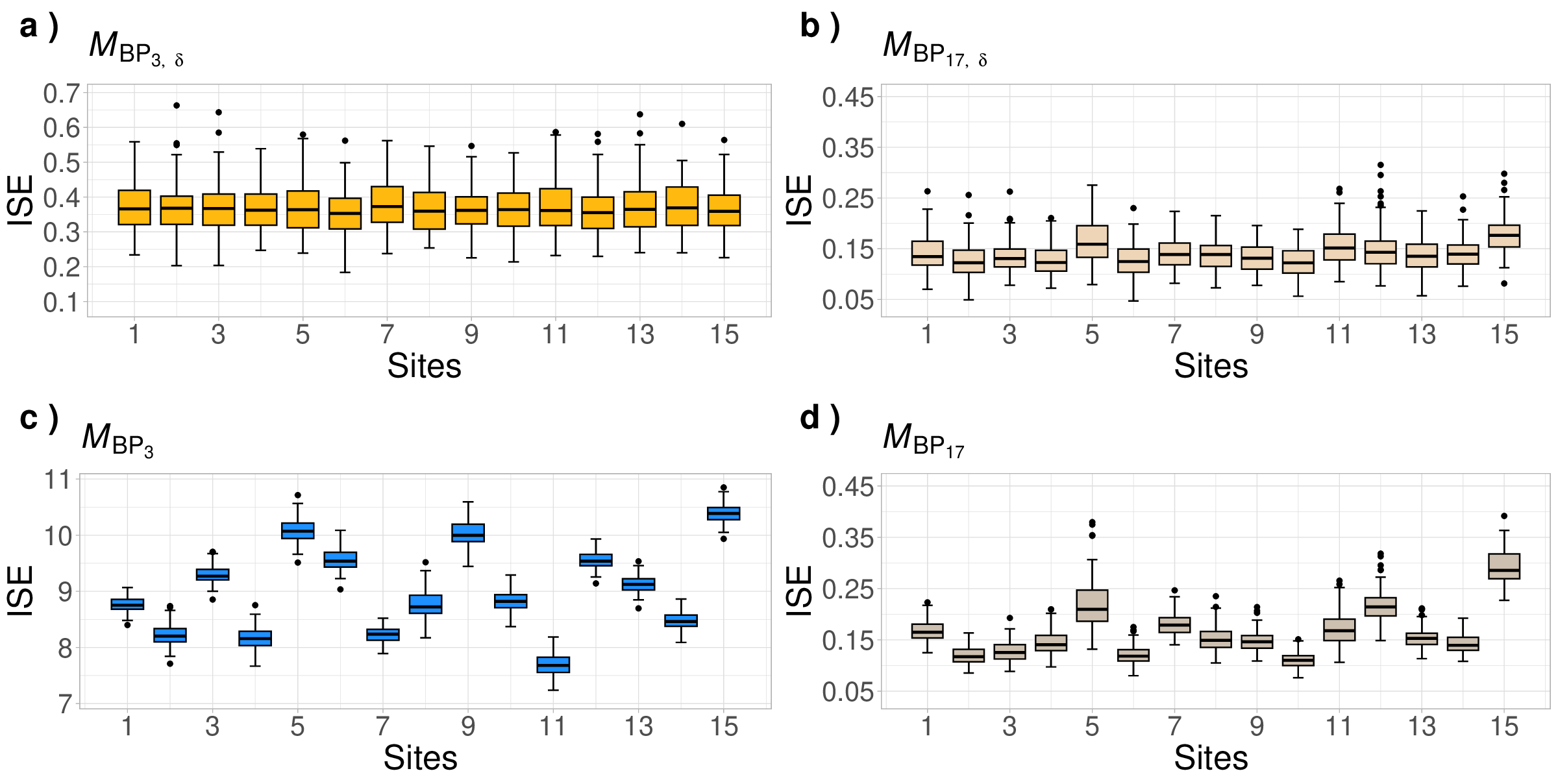}
	\caption{Box plots of the ISEs (100 replicates) are presented to evaluate the estimated curves' performance compared to the target functions. The proposed model fits the data using 4 and 18 BP functional bases, with and without including the random effect component, in Panels (a-d). We consider the values of $\kappa=2$, $\varphi = 1$ and $\sigma^{2} = 2$.}
	\label{fig:3.4}
\end{figure}

In Figure \ref{fig:3.4}, Panel $(a)$ shows that most of the ISE values presented in each curve's boxplot are below 0.4. As the number of bases increases to 18, the median of the majority of the boxplots declines to values below 0.15, as illustrated in panel $(b)$. The variability of the values remains consistently similar across all 15 plots in both panels. This pattern, along with the values obtained from the ISE measurement, indicates the effectiveness of the proposed methodology in successfully recovering curves that exhibit spatial dependencies differing from the model structure $\mathcal{M}_{BP_{p};\,\delta}$. In other words, the smoothed trajectories align closely with the target curves. In contrast, panels $(b)$ and $(c)$ demonstrate that the performance of the $\mathcal{M}_{BP_{p};\,\bullet}$ model is less effective, mainly when using only 4 BP functional bases. This is due to the model's failure to account for potential dependencies between observations within each series, as the $\mathcal{M}_{BP_{p};\,\bullet}$ model does not include a random effect component.

This second part of the study aims to assess how well the model can predict curves in geographical areas that have yet to be observed. To achieve this, we will use sites 1 and 11 from the grid shown in Figure \ref{fig:3.1} as the basis for our predictions. These locations are not part of the sample used to train the model, but we know their actual values, which will help us evaluate the accuracy of the predictions.

\begin{figure}[!hbt]	
	\centering
	\includegraphics[width=1\textwidth]{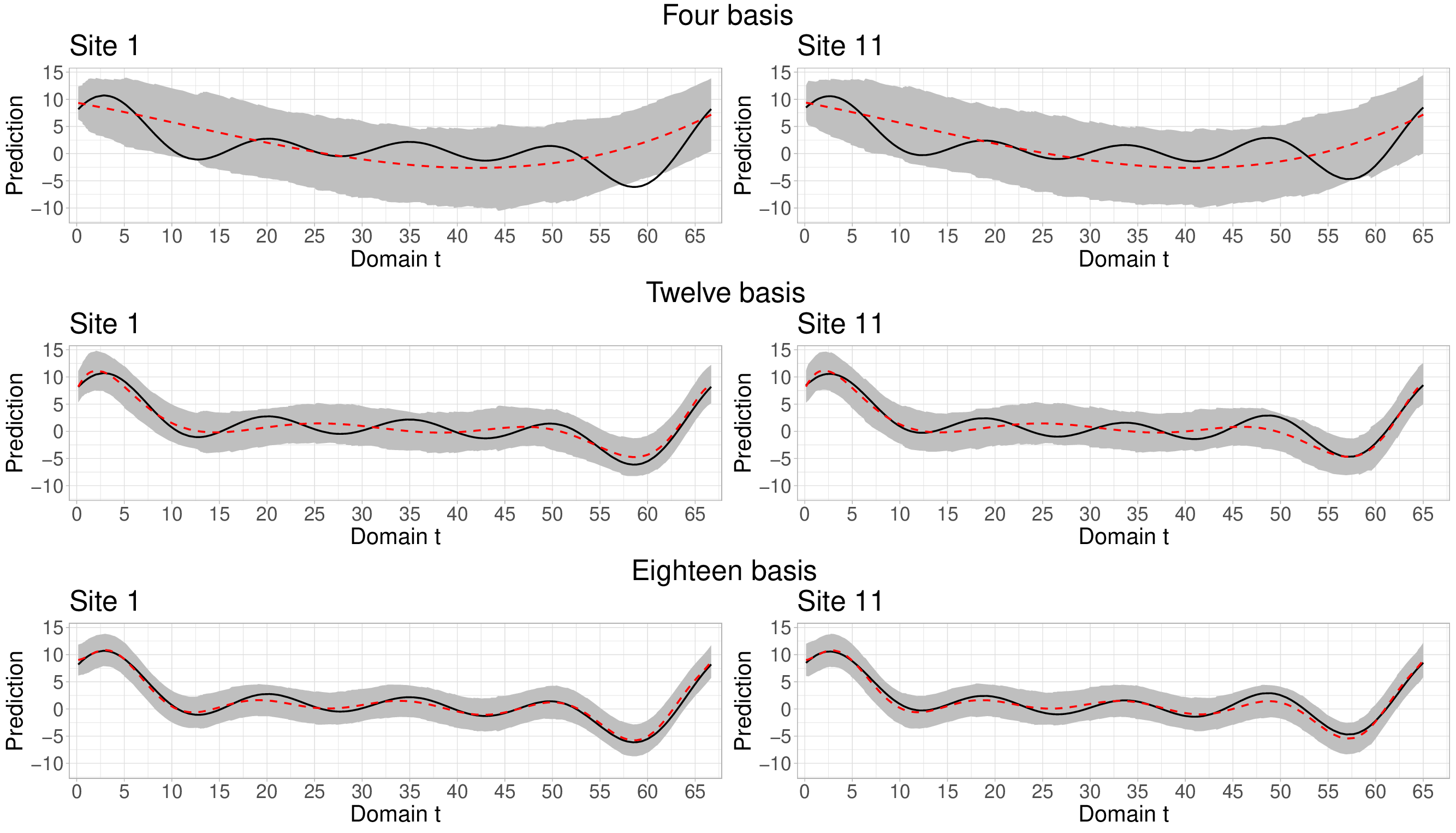}
	\caption{Average (100 replicas) 95\% HPD intervals (shaded area) for predictions. Objective functions (black lines) and average (100 replicas) posterior mean (red and blue dashed lines) for the unobserved sites $s_1$ and $s_{11}$.}
	\label{fig:3.5}
\end{figure}

Figure \ref{fig:3.5} shows the average curve of 100 replicas of the a posteriori means at unobserved locations 1 and 11. These locations are known for having few close neighbors. The observed curves were created under a moderate spatial dependence ($\varphi = 1$). The main goal of the analysis is to assess how well the model predicts these curves using different numbers of bases: 4, 12, and 18. First, it was observed that using a reduced number of basis functions is not enough to accurately capture the spatial structure, which leads to decreased prediction accuracy. However, when 12 BP bases of degree 11 were used, the model showed greater flexibility, resulting in better capture of the trends of the target trajectories. Finally, with 18 functional bases, the smoothed curves closely fit the target curves, demonstrating a remarkable approximation. Additionally, it was observed that, as the number of bases increased, the average (out of 100 replicates) of the 95\% highest posterior density (HPD) intervals (indicated by the gray area) became progressively more compressed, indicating a decrease in uncertainty and increased accuracy in the model predictions.

It should be noted that there is no fixed methodology to determine the optimal number of basis functions. However, increasing their number provides greater flexibility in capturing trends in locations with few neighbors. However, this increase also entails higher computational costs, which can lead to problems related to numerical stability.
\begin{figure}[!hbt]	
	\centering
	\includegraphics[width=1\textwidth]{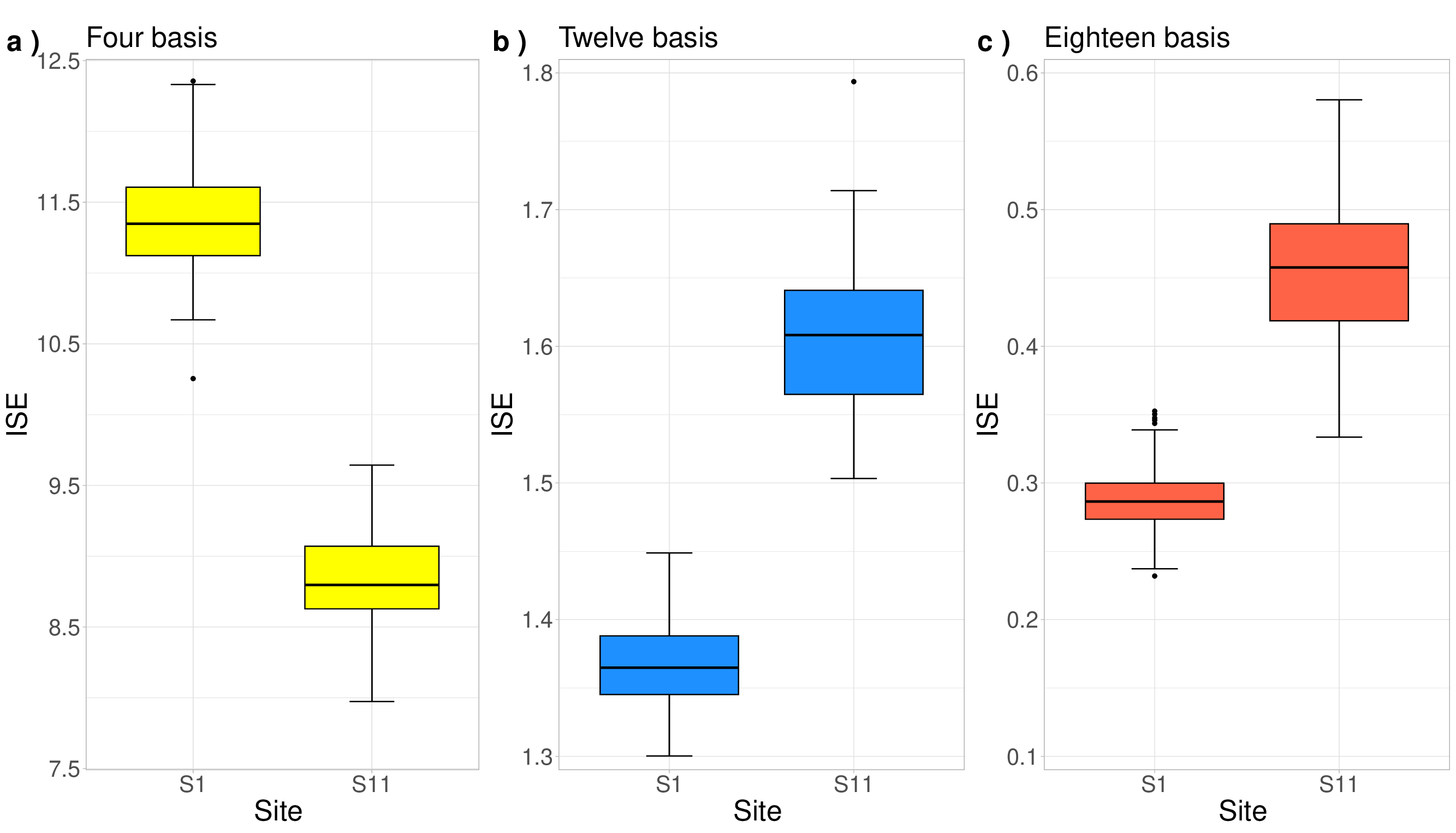}
	\caption{The boxplots illustrate the ISE metric measurements based on 100 replicates. These plots present the results for the unobserved stations, $s_{1}$ and $s_{11}$. They assess how closely the smoothed curves align with the target functions by employing the proposed model to fit the data using four BP functional bases in Panel (a), twelve bases in Panel (b), and eighteen bases in Panel (c).}
	\label{fig:3.6}
\end{figure}

Figure \ref{fig:3.6} presents the boxplots for 100 measurements of the ISE metric, with one measurement for each replicate, which allows us to evaluate the proximity of the smoothed curves concerning the target functions. Firstly, when a reduced number of basis functions is used (Panel a),  the prediction curve at location \(s_{11}\), which is more isolated from neighboring locations, demonstrates better performance compared to \(s_{1}\). However, increasing the number of basis functions enhances the model's flexibility, allowing it to capture spatial dependence better and improve estimates at both locations, with \(s_{1}\) exhibiting superior performance. However, this increased flexibility also results in more significant variability in the ISE values, especially in the $s_{11}$ boxplot (Panel b and c). Finally, as the number of basis functions increases, the ISE values decrease, indicating an improvement in overall performance. This effect is especially noticeable in $s_{1}$, whose proximity to other locations strengthens the inference process and improves smoothing. The Appendix \ref{app:B} offers supplementary details on other scenarios explored in this study.

\section{Real Data Application}\label{Cap:4}
Particulate matter, commonly called PM$_{10}$, consists of tiny particles suspended in the air. These particles have a diameter of 10 micrometers or less, making them small enough to be inhaled. This characteristic poses a significant risk to human health as it can penetrate deep into our respiratory system, causing damage to tissues and organs. Moreover, PM$_{10}$ can serve as a carrier for bacteria and viruses, potentially exacerbating the spread of diseases. Extensive research has established a positive relationship between exposure to PM$_{10}$ and various adverse health outcomes. Studies by \cite{greenbaum2001} and \cite{paldy2006} have highlighted the association between PM$_{10}$ exposure and an increased risk of respiratory and cardiovascular diseases, cancer, influenza, and asthma. These findings underscore the importance of monitoring and mitigating the levels of particulate matter in the air we breathe.

The Mexican Official Standard, a NOM-025-SSA1-2021 (Norma Oficial Mexicana), establishes the concentration limits for suspended particulate matter (PM$_{10}$) in ambient air to protect public health. This standard outlines the criteria for assessing these concentration levels. Specifically, it sets a 24-hour average limit for acute exposure at 70 $\mu\text{g/m}^{3}$ (which corresponds to a $\log(\text{PM}_{10})$ scale value of 4.3) and an average annual limit for chronic exposure of 36 $\mu\text{g/m}^{3}$ (with a $\log(\text{PM}_{10})$ value of 3.6). These values are significantly higher than the World Health Organization (WHO) air quality guidelines, which recommend a limit of less than 20 $\mu\text{g/m}^{3}$ ($\log(\text{PM}_{10})$ value of 3).

\begin{figure}[!hbt]	
	\centering
	\includegraphics[width=1\textwidth]{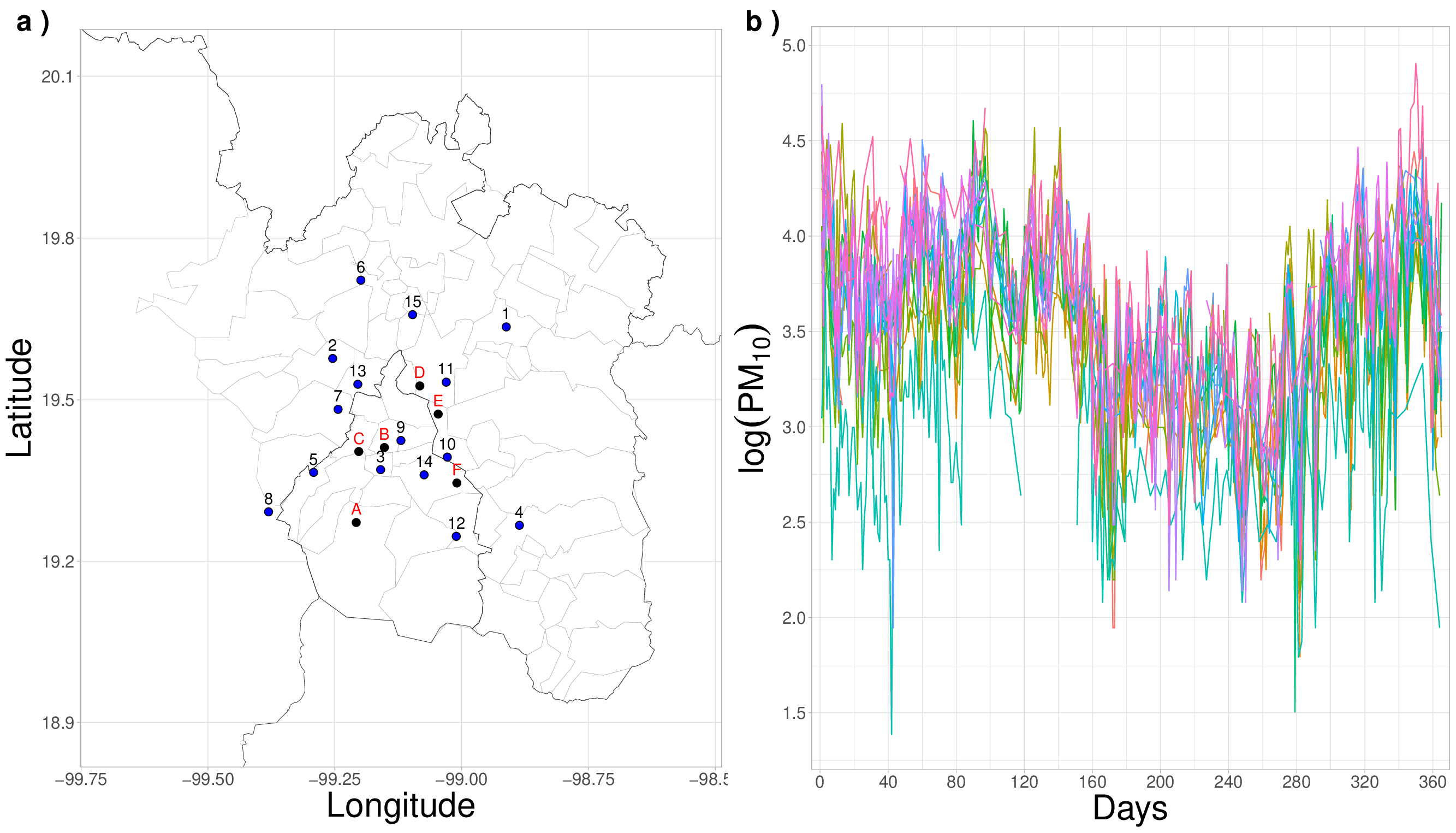}
	\caption{Map and data. Panel (a) shows the map of Mexico City indicating the location of 15 active and six inactive monitoring stations. Panel (b) presents the PM$_{10}$ data recorded during the year 2022.}
	\label{fig:3.7}
\end{figure}

For one year, data on air quality in Mexico City was collected through a monitoring network. This data corresponds to consecutive hours, from 1:00 am. on January 1, 2022, to midnight on December 31, 2022. Monitoring was conducted at 17 environmental stations operating throughout the year and at four others out of service during the study period. These stations are located at different points in the city, as illustrated in Figure \ref{fig:3.7} Panel($a$). They are all part of the Automatic Atmospheric Monitoring Network (RAMA) and measure, among other parameters, particulate matter down to 10 micrometers ($mu$) every hour. The data collected are publicly available on a web page managed by the Mexico City government.

Observations are gathered at discrete, evenly spaced time intervals (measured in hours), but some observations are missing values. This study examines a scenario with irregular time intervals, requiring adjustments to reflect this aspect in the PM$_{10}$ data. The following steps outline the necessary actions to address this characteristic: 
\begin{itemize}
    \item[i.] A separate analysis is conducted for each month of the study period 2022 to calculate the percentage of missing data per month for each curve individually. This approach allows each curve to reflect different patterns of irregular spacing. The PM$_{10}$ dataset covers 15 stations, each with 8,760 hours of recorded data. 
    \item[ii.] Quintiles are calculated to gather information on the dispersion of missing data. This analysis provides categories identifying the months with the least and most missing data in each curve. A spacing pattern is then established for the observations of each month, as shown in Table \ref{tb:4.3}. For example, if a month has between 40.1\% and 60\% missing values, the analysis considers a spacing that summarizes information corresponding to 48 hours (equivalent to 2 days).
\end{itemize}

\begin{table}[!h]
 \centering
  \caption{Categories for classifying months based on the amount of missing data for each curve: Column 1 shows the range for each category, Column 2 specifies the number of hours used to summarize the response, and Column 3 indicates the equivalent number of days.} \label{tb:4.3}
{\scalebox{1}{ \begin{tabular}{ccc}
\hline
Category for & Number of hours & Domain \\
the month    & to summarize    & in days\\
\hline
$0\%-20\%$     & 24  & 1\\[0.15cm]
$20.1\%-40\%$  & 24  & 1\\[0.15cm]
$40.1\%-60\%$  & 48  & 2\\[0.15cm]
$60.1\%-80\%$  & 72  & 3\\[0.15cm] 
$80.1\%-100\%$ & 120 & 5\\[0.15cm] 
\hline
\end{tabular}}}
\end{table}
\begin{itemize}
    \item[iii.] We calculate the median of the data recorded at intervals of 24, 48, 72, and 120 hours according to the previous configuration, which makes it easier to work with a day-based functional domain. The first day of each interval, in which the median is obtained, is the time $t_{ij}$ in the functional domain. If the interval contains only missing observations, the location $t_{ij}$ is considered a point with a missing value, which must be estimated. Notably, missing values do not hinder the proposed model, as it can effectively handle them, as demonstrated in Appendix \ref{app:B}.
\end{itemize}
After following these steps, the resulting samples exhibit series with distinct patterns of irregular spacing between discrete points in the domain and different numbers of observations, spanning both PM$_{10}$ values and missing data. For more details, see Table \ref{tb:4.4}.
\begin{table}[!h]
 \centering
  \caption{Number of valid (non-NA) and missing observations for each of the 15 curves in the observed sample.} \label{tb:4.4}
{\scalebox{1}{ \begin{tabular}{ccccccccc}
\hline
&\multicolumn{2}{c}{Observations}&&\multicolumn{2}{c}{Observations}&&\multicolumn{2}{c}{Observations}\\
\cline{2-3} \cline{5-6} \cline{8-9}
Site & Non-NA  & NA  & Site & Non-NA  & NA  & Site & Non-NA  & NA \\
\hline
1  & 210  & 11  & 6  & 218  & 4   & 11 & 214  & 8    \\[0.15cm]
2  & 209  & 14  & 7  & 216  & 8   & 12 & 209  & 12   \\[0.15cm] 
3  & 226  & 1   & 8  & 215  & 6   & 13 & 212  & 10   \\[0.15cm]
4  & 210  & 13  & 9  & 217  & 5   & 14 & 214  & 8    \\[0.15cm]
5  & 213  & 8   & 10 & 223  & 3   & 15 & 216  & 7    \\[0.15cm]
\hline
\end{tabular}}}
\end{table}

During the study period 2022, stations A, B, C, D, E, and F were not operational, which impacted data collection, see Figure \ref{fig:3.7} Panel (a). The primary goal of this study is to forecast the variations in PM$_{10}$ concentration over time, utilizing 250 distinct time points that span the 365 days of the year. The distribution of these time points adheres to a spacing pattern that mirrors the configuration used in the reference sample set.

\begin{figure}[!hbt]	
	\centering
	\includegraphics[width=1\textwidth]{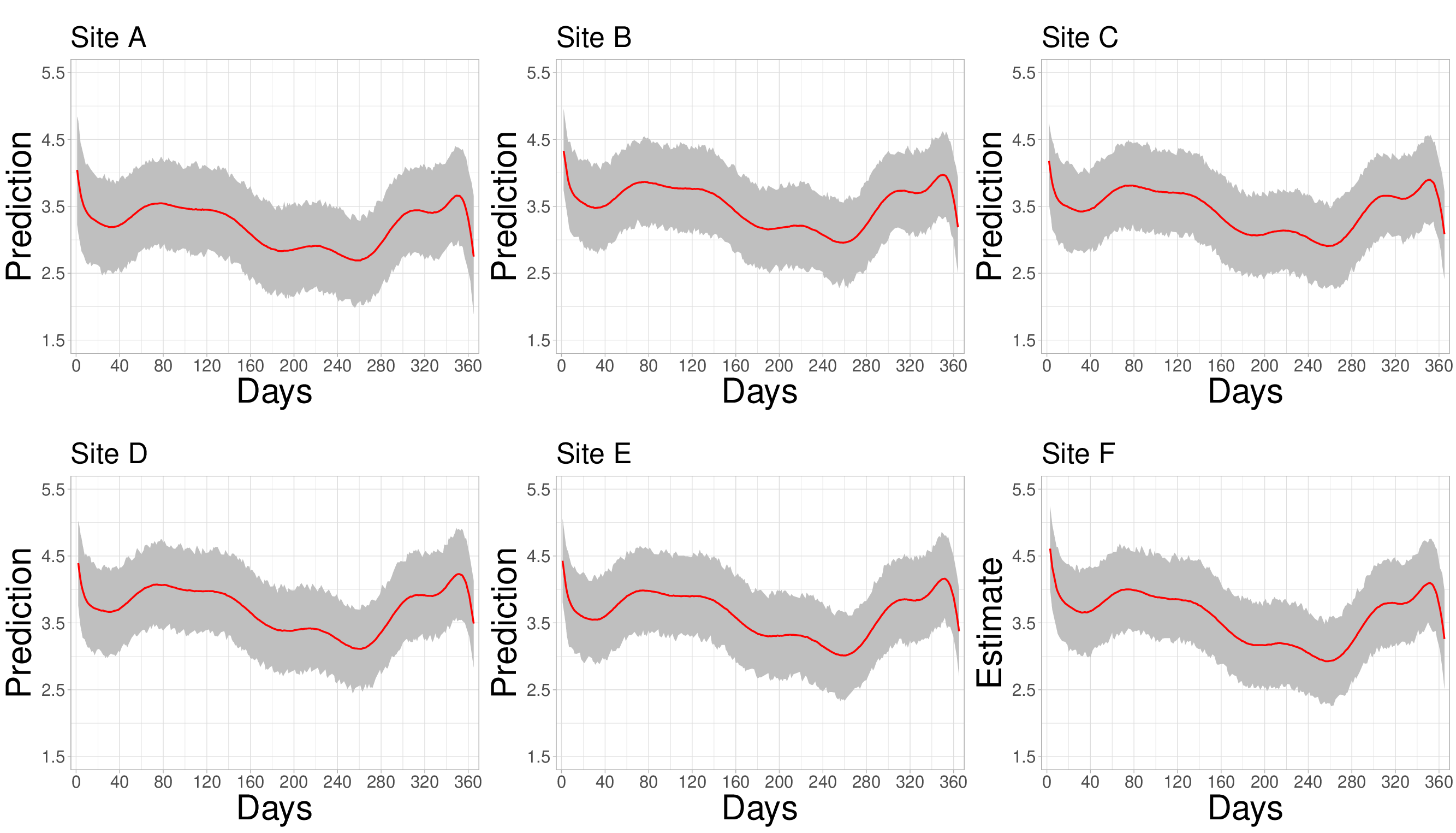}
	\caption{The prediction graphs show PM$_{10}$ values for six locations (Site A to Site F) over several days in 2022. The red line represents the predicted trends for PM$_{10}$ concentrations, while the shaded area indicates the range of uncertainty around these estimates, known as the 95\% HPD intervals.}
	\label{fig:3.8}
\end{figure}

In Figure \ref{fig:3.8}, the Bayesian hierarchical model, which employs 18 BP functional bases, demonstrates its predictive capability for PM$_{10}$ concentrations at six monitoring stations in Mexico City, which were not operational throughout the year. The red line represents the predicted PM$_{10}$ trends, while the shaded area denotes the 95\% HPD intervals, capturing prediction uncertainty. The figure suggests reliability in the predictions, as evidenced by the variation in HPD interval widths across sites. Locations with narrower uncertainty bands indicate more robust and consistent predictions, likely influenced by data from nearby operational stations with similar environmental conditions (see site B panel). Conversely, broader intervals suggest increased uncertainty, particularly in isolated sites with fewer geographic neighbors (see site A panel).  

During almost the entire period analyzed, five out of six inactive air quality monitoring stations recorded PM$_{10}$ concentration levels that consistently exceeded the limits established by national regulations and the WHO. In contrast, the curve for station A displayed notably different behavior; its concentration remained below 3.6, the limit set by official Mexican standards, for most of the assessment period. A possible explanation for this behavior lies in the geographical characteristics of the area surrounding station A, which corresponds to an isolated neighborhood with low vehicular flow, little industrial activity, and green regions that help mitigate pollution.

Applying a Bayesian hierarchical model to PM$_{10}$ concentrations in Mexico City has shown its ability to provide accurate predictions for areas lacking active monitoring stations. The results emphasize the model's flexibility in handling irregularly spaced observations and its effectiveness in estimating temporal trends and spatial dependencies. This predictive capability is essential for formulating public health policies, as it allows authorities to identify areas at a higher risk of pollution-related health impacts, even without direct measurements. By integrating these predictions with geographical and environmental factors, policymakers can better prioritize interventions, allocate resources efficiently, and develop preventive measures tailored to the specific needs of under-monitored regions. This approach improves air quality management and highlights the importance of maintaining and expanding monitoring networks to support evidence-based health policy decisions.

\section{Conclusions.}\label{Cap:5}
This study presents an essential advancement in Bayesian functional models by incorporating a spatial dependence structure that efficiently addresses the irregular spacing of observations. The proposed methodology employs Bernstein polynomial basis functions and an autoregressive random-effects component. This approach allows us to accurately capture the spatial correlation and dependencies, whether temporal or related to any continuous variable, linked to the irregular distribution of the discrete points associated with the observations. In this way, the model's ability to analyze and interpret complex data patterns, typical of real-world scenarios, is substantially improved. Additionally, since polynomials are employed, it is important to note that this procedure can also effectively capture the dynamics of concentration curves, similar to the findings of \cite{Colla2023}. 

The results of the simulations indicated that the model is effective in recovering curves with spatial dependence y in situations of irregular spacing, maintaining low and consistent levels of ISE. In addition, the prediction analysis for unobserved locations showed that the model could significantly capture the spatial and functional variability of the curves when increasing the number of basis functions. In its application to PM$_{10}$ concentration data, the model demonstrated its ability to predict time series in locations where the monitoring stations were not operational, generating estimates with HPD intervals that reflect the degree of uncertainty associated with the spatial variation of the data.

 
 Our modeling framework introduces significant differences and enhancements over the approach proposed by \cite{Burbano-Moreno:2024a, Burbano-Moreno:2024b}, particularly in effectively addressing the irregular spacing of discrete points within the functional domain. The model proposed by them employs a standard Gaussian cumulative function to handle irregular distances, where the argument of this function depends on the prior knowledge that the researcher has about the data set. This approach requires a transformation of the data into the interval (0,1), which allows control of the association between consecutive random effects according to the positions $t_{i}$ and $t_{i-1}$. In contrast, this approach employs a first-order autoregressive process for the variable $\delta$, which links the irregular spacing without the need to transform the distances. In this case, the autoregressive parameter $\phi$ follows an exponential structure, with a parameter $\eta$ that controls the relationship between successive random effects. This parameter is unknown, so a noninformative prior distribution is assigned, allowing the data to estimate it adequately. This choice improves the model's ability to fit temporal structures and other highly variable continuous domains, contributing to higher predictive accuracy.

There are several potential future directions for improving this model. One development area could involve incorporating more complex spatial and functional variability structures, particularly in situations with high dimensionality or significant non-linear temporal dependencies. Another promising direction would be to investigate alternative methods for adaptively choosing the number of basis functions to balance model flexibility and numerical stability, especially in contexts with high volumes of data or highly irregular curves. There is also significant potential in applying the model to environmental and health-related time series data, which could lead to advancements in functional and spatial analysis methodologies. Furthermore, when applied to similar real-world data, this new procedure may help policymakers and epidemiologists make appropriate decisions regarding public health.

\bibliographystyle{chicago}
\bibliography{References}

\begin{thebibliography}{}

\bibitem[\protect\citeauthoryear{Aguilera-Morillo, Durb{\'a}n, and Aguilera}{Aguilera-Morillo et~al.}{2017}]{aguilera2017}
Aguilera-Morillo, M.~C., M.~Durb{\'a}n, and A.~M. Aguilera (2017).
\newblock Prediction of functional data with spatial dependence: a penalized approach.
\newblock {\em Stochastic Environmental Research and Risk Assessment\/}~{\em 31\/}(1), 7--22.

\bibitem[\protect\citeauthoryear{Baladandayuthapani, Mallick, Young-Hong, Lupton, Turner, and Carroll}{Baladandayuthapani et~al.}{2008}]{baladandayuthapani:2008}
Baladandayuthapani, V., B.~K. Mallick, M.~Young-Hong, J.~R. Lupton, N.~D. Turner, and R.~J. Carroll (2008).
\newblock Bayesian hierarchical spatially correlated functional data analysis with application to colon carcinogenesis.
\newblock {\em Biometrics\/}~{\em 64\/}(1), 64--73.

\bibitem[\protect\citeauthoryear{Banerjee, Carlin, and Gelfand}{Banerjee et~al.}{2014}]{banerjee:2014}
Banerjee, S., B.~P. Carlin, and A.~E. Gelfand (2014).
\newblock {\em Hierarchical Modeling and Analysis for Spatial Data\/} (2 ed.).
\newblock London: Chapman and Hall/CRC.

\bibitem[\protect\citeauthoryear{Bernstein}{Bernstein}{1912}]{bernvstein:1912}
Bernstein, S. (1912).
\newblock {D{\'e}monstration du Th{\'e}oreme de Weierstrass Fond{\'e}e Sur le Calcul des Probabilities}.
\newblock {\em Communications of the Kharkov Mathematical\/}~{\em 13}, 1--2.

\bibitem[\protect\citeauthoryear{Brown, Le, and Zidek}{Brown et~al.}{1994}]{brown:1994}
Brown, P.~J., N.~D. Le, and J.~V. Zidek (1994).
\newblock Multivariate spatial interpolation and exposure to air pollutants.
\newblock {\em Canadian Journal of Statistics\/}~{\em 22\/}(4), 489--509.

\bibitem[\protect\citeauthoryear{Burbano-Moreno and Mayrink}{Burbano-Moreno and Mayrink}{2024a}]{Burbano-Moreno:2024b}
Burbano-Moreno, A.~A. and V.~D. Mayrink (2024a).
\newblock Gaussian modeling with b-splines for spatial functional data on irregular domains.
\newblock {\em Statistics\/}~{\em 58\/}(6), 1304--1331.

\bibitem[\protect\citeauthoryear{Burbano-Moreno and Mayrink}{Burbano-Moreno and Mayrink}{2024b}]{Burbano-Moreno:2024a}
Burbano-Moreno, A.~A. and V.~D. Mayrink (2024b).
\newblock Spatial functional data analysis: Irregular spacing and bernstein polynomials.
\newblock {\em Spatial Statistics\/}~{\em 60}, 100832.

\bibitem[\protect\citeauthoryear{Collazos, Dias, and Medeiros}{Collazos et~al.}{2023}]{Colla2023}
Collazos, J. A.~A., R.~Dias, and M.~C. Medeiros (2023).
\newblock Modeling the evolution of deaths from infectious diseases with functional data models: The case of covid-19 in brazil.
\newblock {\em Statistics in Medicine\/}~{\em 42\/}(7), 993--1012.

\bibitem[\protect\citeauthoryear{Cressie}{Cressie}{1993}]{cressie:1993}
Cressie, N. (1993).
\newblock {\em Statistics for Spatial Data}.
\newblock New York: John Wiley and Sons.

\bibitem[\protect\citeauthoryear{Datta, Banerjee, Finley, and Gelfand}{Datta et~al.}{2016}]{datta:2016}
Datta, A., S.~Banerjee, A.~O. Finley, and A.~E. Gelfand (2016).
\newblock Hierarchical nearest-neighbor {G}aussian process models for large geostatistical datasets.
\newblock {\em Journal of the American Statistical Association\/}~{\em 111\/}(514), 800--812.

\bibitem[\protect\citeauthoryear{Davis}{Davis}{1975}]{davis:1975}
Davis, P. (1975).
\newblock {\em Interpolation and Approximation}.
\newblock New York: Dover Publications.

\bibitem[\protect\citeauthoryear{De-Boor}{De-Boor}{2001}]{boor:2001}
De-Boor, C. (2001).
\newblock {\em A Practical Guide to Splines}.
\newblock New York: Springer.

\bibitem[\protect\citeauthoryear{Diggle and Ribeiro}{Diggle and Ribeiro}{2007}]{diggle:2007}
Diggle, P.~J. and P.~J. Ribeiro (2007).
\newblock {\em Model-Based Geostatistics}.
\newblock New York: Springer.

\bibitem[\protect\citeauthoryear{Farouki and Rajan}{Farouki and Rajan}{1987}]{farouki:1987}
Farouki, R.~T. and V.~T. Rajan (1987).
\newblock On the numerical condition of polynomials in {B}ernstein form.
\newblock {\em Computer Aided Geometric Design\/}~{\em 4\/}(3), 191--216.

\bibitem[\protect\citeauthoryear{Farouki and Rajan}{Farouki and Rajan}{1988}]{farouki:1988}
Farouki, R.~T. and V.~T. Rajan (1988).
\newblock Algorithms for polynomials in {B}ernstein form.
\newblock {\em Computer Aided Geometric Design\/}~{\em 5\/}(1), 1--26.

\bibitem[\protect\citeauthoryear{Ferraty and Vieu}{Ferraty and Vieu}{2006}]{ferraty:2006}
Ferraty, F. and P.~Vieu (2006).
\newblock {\em Nonparametric Functional Data Analysis: Theory and Practice}.
\newblock New York: Springer.

\bibitem[\protect\citeauthoryear{Fu and Heckman}{Fu and Heckman}{2019}]{fu:2019}
Fu, E. and N.~Heckman (2019).
\newblock Model-based curve registration via stochastic approximation em algorithm.
\newblock {\em Computational Statistics \& Data Analysis\/}~{\em 131}, 159--175.

\bibitem[\protect\citeauthoryear{Gelfand, Banerjee, and Gamerman}{Gelfand et~al.}{2005}]{gelfand:2005}
Gelfand, A.~E., S.~Banerjee, and D.~Gamerman (2005).
\newblock Spatial process modelling for univariate and multivariate dynamic spatial data.
\newblock {\em Environmetrics\/}~{\em 16\/}(5), 465--479.

\bibitem[\protect\citeauthoryear{Gentle}{Gentle}{2009}]{gentle:2009}
Gentle, J. (2009).
\newblock {\em Computational Statistics}.
\newblock New York: Springer.

\bibitem[\protect\citeauthoryear{Giraldo, Delicado, and Mateu}{Giraldo et~al.}{2012}]{giraldo:2012}
Giraldo, R., P.~Delicado, and J.~Mateu (2012).
\newblock {Hierarchical Clustering of Spatially Correlated Functional Data}.
\newblock {\em Statistica Neerlandica\/}~{\em 66\/}(4), 403--421.

\bibitem[\protect\citeauthoryear{Greenbaum, Bachmann, Krewski, Samet, White, and Wyzga}{Greenbaum et~al.}{2001}]{greenbaum2001}
Greenbaum, D.~S., J.~D. Bachmann, D.~Krewski, J.~M. Samet, R.~White, and R.~E. Wyzga (2001).
\newblock Particulate air pollution standards and morbidity and mortality: Case study.
\newblock {\em American Journal of Epidemiology\/}~{\em 154\/}(12), S78--S90.

\bibitem[\protect\citeauthoryear{Guo, Kurtek, and Bharath}{Guo et~al.}{2022}]{guo:2022}
Guo, X., S.~Kurtek, and K.~Bharath (2022).
\newblock Variograms for kriging and clustering of spatial functional data with phase variation.
\newblock {\em Spatial Statistics\/}~{\em 51}, 100687.

\bibitem[\protect\citeauthoryear{Hoffman and Gelman}{Hoffman and Gelman}{2014}]{Hoffman2014}
Hoffman, M.~D. and A.~Gelman (2014).
\newblock The {No-U-Turn} sampler: Adaptively setting path lengths in {H}amiltonian {M}onte {C}arlo.
\newblock {\em Journal of Machine Learning Research\/}~{\em 15\/}(1), 1593--1623.

\bibitem[\protect\citeauthoryear{Hollander, Wolfe, and Chicken}{Hollander et~al.}{2013}]{hollander:2013}
Hollander, M., D.~A. Wolfe, and E.~Chicken (2013).
\newblock {\em Nonparametric Statistical Methods}.
\newblock New Jersey: Wiley.

\bibitem[\protect\citeauthoryear{James}{James}{2010}]{Gareth:2010}
James, G. (2010, 11).
\newblock {Sparseness and Functional Data Analysis}.
\newblock In {\em {The Oxford Handbook of Functional Data Analysis}}. Oxford University Press.

\bibitem[\protect\citeauthoryear{James, Hastie, and Sugar}{James et~al.}{2000}]{james:2000}
James, G.~M., T.~J. Hastie, and C.~A. Sugar (2000).
\newblock Principal component models for sparse functional data.
\newblock {\em Biometrika\/}~{\em 87\/}(3), 587--602.

\bibitem[\protect\citeauthoryear{Jiang and Serban}{Jiang and Serban}{2012}]{jiang:2012}
Jiang, H. and N.~Serban (2012).
\newblock Clustering random curves under spatial interdependence with application to service accessibility.
\newblock {\em Technometrics\/}~{\em 54\/}(2), 108--119.

\bibitem[\protect\citeauthoryear{Kim, Park, and Kim}{Kim et~al.}{2021}]{kim:2021}
Kim, S.-H., J.-W. Park, and J.-H. Kim (2021).
\newblock Functional data analysis for assessing the fatigue life of construction equipment attachments.
\newblock {\em Journal of Mechanical Science and Technology\/}~{\em 35}, 495--506.

\bibitem[\protect\citeauthoryear{Kokoszka and Reimherr}{Kokoszka and Reimherr}{2017}]{kokoszka:2017}
Kokoszka, P. and M.~Reimherr (2017).
\newblock {\em Introduction to Functional Data Analysis}.
\newblock London: Chapman and Hall/CRC.

\bibitem[\protect\citeauthoryear{Korte-Stapff, Yarger, Stoev, and Hsing}{Korte-Stapff et~al.}{2022}]{korte:2022}
Korte-Stapff, M., D.~Yarger, S.~Stoev, and T.~Hsing (2022).
\newblock A multivariate functional-data mixture model for spatio-temporal data: Inference and cokriging.
\newblock {\em arXiv preprint arXiv:2211.04012\/}.

\bibitem[\protect\citeauthoryear{Lawson}{Lawson}{2021}]{lawson:2021}
Lawson, A.~B. (2021).
\newblock {\em Using {R} for {B}ayesian Spatial and Spatio-Temporal Health Modeling}.
\newblock New York: CRC Press.

\bibitem[\protect\citeauthoryear{Li and Luo}{Li and Luo}{2017}]{li:2017}
Li, K. and S.~Luo (2017).
\newblock Functional joint model for longitudinal and time-to-event data: an application to alzheimer's disease.
\newblock {\em Statistics in Medicine\/}~{\em 36\/}(22), 3560--3572.

\bibitem[\protect\citeauthoryear{Liu, Ray, and Hooker}{Liu et~al.}{2017}]{liu2017}
Liu, C., S.~Ray, and G.~Hooker (2017).
\newblock Functional principal component analysis of spatially correlated data.
\newblock {\em Statistics and Computing\/}~{\em 27\/}(6), 1639--1654.

\bibitem[\protect\citeauthoryear{Liu and Guo}{Liu and Guo}{2012}]{liu:2012}
Liu, Z. and W.~Guo (2012).
\newblock Functional mixed effects models.
\newblock {\em Wiley Interdisciplinary Reviews: Computational Statistics\/}~{\em 4\/}(6), 527--534.

\bibitem[\protect\citeauthoryear{Lorentz}{Lorentz}{2012}]{lorentz:1986}
Lorentz, G.~G. (2012).
\newblock {\em Bernstein Polynomials\/} (2 ed.).
\newblock New York: American Mathematical Society.

\bibitem[\protect\citeauthoryear{Mart{\'i}nez-Hern{\'a}ndez and Genton}{Mart{\'i}nez-Hern{\'a}ndez and Genton}{2020}]{Israel:2020}
Mart{\'i}nez-Hern{\'a}ndez, I. and M.~G. Genton (2020).
\newblock {Recent Developments in Complex and Spatially Correlated Functional Data}.
\newblock {\em Brazilian Journal of Probability and Statistics\/}~{\em 34\/}(2), 204--229.

\bibitem[\protect\citeauthoryear{Mateu and Giraldo}{Mateu and Giraldo}{2021}]{mateu:2021}
Mateu, J. and R.~Giraldo (2021).
\newblock {\em Geostatistical Functional Data Analysis}.
\newblock Hoboken: John Wiley and Sons.

\bibitem[\protect\citeauthoryear{Mateu and Romano}{Mateu and Romano}{2017}]{Jorge:2017}
Mateu, J. and E.~Romano (2017).
\newblock Advances in spatial functional statistics.
\newblock {\em Stochastic Environmental Research and Risk Assessment\/}~{\em 31}, 1--6.

\bibitem[\protect\citeauthoryear{Morris and Carroll}{Morris and Carroll}{2006}]{morris:2006}
Morris, J.~S. and R.~J. Carroll (2006).
\newblock Wavelet-based functional mixed models.
\newblock {\em Journal of the Royal Statistical Society Series B: Statistical Methodology\/}~{\em 68\/}(2), 179--199.

\bibitem[\protect\citeauthoryear{M{\"u}ller, Sen, and Stadtm{\"u}ller}{M{\"u}ller et~al.}{2011}]{muller:2011}
M{\"u}ller, H.-G., R.~Sen, and U.~Stadtm{\"u}ller (2011).
\newblock Functional data analysis for volatility.
\newblock {\em Journal of Econometrics\/}~{\em 165\/}(2), 233--245.

\bibitem[\protect\citeauthoryear{Nerini, Monestiez, and Mant{\'e}}{Nerini et~al.}{2010}]{Nerini:2010}
Nerini, D., P.~Monestiez, and C.~Mant{\'e} (2010).
\newblock {Cokriging for Spatial Functional Data}.
\newblock {\em Journal of Multivariate Analysis\/}~{\em 101\/}(2), 409--418.

\bibitem[\protect\citeauthoryear{Paldy, Bobvos, Lustigova, Moshammer, Niciu, Otorepec, Puklova, Szafraniec, Zagargale, Neuberger, et~al.}{Paldy et~al.}{2006}]{paldy2006}
Paldy, A., J.~Bobvos, M.~Lustigova, H.~Moshammer, E.~M. Niciu, P.~Otorepec, V.~Puklova, K.~Szafraniec, T.~Zagargale, M.~Neuberger, et~al. (2006).
\newblock Health impact assessment of pm10 on mortality and morbidity in children in central-eastern european cities.
\newblock {\em Epidemiology\/}~{\em 17\/}(6), S131.

\bibitem[\protect\citeauthoryear{{R Core Team}}{{R Core Team}}{2023}]{R:2023}
{R Core Team} (2023).
\newblock {\em R: A Language and Environment for Statistical Computing}.
\newblock Vienna, Austria: R Foundation for Statistical Computing.

\bibitem[\protect\citeauthoryear{Ramsay and Silverman}{Ramsay and Silverman}{2005}]{ramsay:2005}
Ramsay, J. and B.~W. Silverman (2005).
\newblock {\em Functional Data Analysis}.
\newblock New York: Springer.

\bibitem[\protect\citeauthoryear{Ramsay and Silverman}{Ramsay and Silverman}{2002}]{ramsay:2002}
Ramsay, J.~O. and B.~W. Silverman (2002).
\newblock {\em Applied Functional Data Analysis: Methods and Case Studies}.
\newblock New York: Springer.

\bibitem[\protect\citeauthoryear{Rekabdarkolaee, Krut, Fuentes, and Reich}{Rekabdarkolaee et~al.}{2019}]{rekabdarkolaee:2019}
Rekabdarkolaee, H.~M., C.~Krut, M.~Fuentes, and B.~J. Reich (2019).
\newblock A {B}ayesian multivariate functional model with spatially varying coefficient approach for modeling hurricane track data.
\newblock {\em Spatial Statistics\/}~{\em 29}, 351--365.

\bibitem[\protect\citeauthoryear{Romano, Balzanella, and Verde}{Romano et~al.}{2017}]{romano:2017}
Romano, E., A.~Balzanella, and R.~Verde (2017).
\newblock Spatial variability clustering for spatially dependent functional data.
\newblock {\em Statistics and Computing\/}~{\em 27}, 645--658.

\bibitem[\protect\citeauthoryear{Shi, Ma, Faisal~Beg, and Cao}{Shi et~al.}{2022}]{shi:2022}
Shi, H., D.~Ma, M.~Faisal~Beg, and J.~Cao (2022).
\newblock A functional proportional hazard cure rate model for interval-censored data.
\newblock {\em Statistical Methods in Medical Research\/}~{\em 31\/}(1), 154--168.

\bibitem[\protect\citeauthoryear{Song and Mallick}{Song and Mallick}{2019}]{song:2019}
Song, J.~J. and B.~Mallick (2019).
\newblock Hierarchical bayesian models for predicting spatially correlated curves.
\newblock {\em Statistics\/}~{\em 53\/}(1), 196--209.

\bibitem[\protect\citeauthoryear{Staicu, Crainiceanu, and Carroll}{Staicu et~al.}{2010}]{staicu:2010}
Staicu, A.-M., C.~M. Crainiceanu, and R.~J. Carroll (2010).
\newblock Fast methods for spatially correlated multilevel functional data.
\newblock {\em Biostatistics\/}~{\em 11\/}(2), 177--194.

\bibitem[\protect\citeauthoryear{{Stan Development Team}}{{Stan Development Team}}{2023}]{stan:2023}
{Stan Development Team} (2023).
\newblock {\em Stan Modeling Language Users Guide and Reference Manual}.
\newblock version 2.18.0.

\bibitem[\protect\citeauthoryear{Thompson and Rosen}{Thompson and Rosen}{2008}]{thompson:2008}
Thompson, W.~K. and O.~Rosen (2008).
\newblock A bayesian model for sparse functional data.
\newblock {\em Biometrics\/}~{\em 64\/}(1), 54--63.

\bibitem[\protect\citeauthoryear{Ver-Hoef and Barry}{Ver-Hoef and Barry}{1998}]{ver:1998}
Ver-Hoef, J.~M. and R.~P. Barry (1998).
\newblock Constructing and fitting models for cokriging and multivariable spatial prediction.
\newblock {\em Journal of Statistical Planning and Inference\/}~{\em 69\/}(2), 275--294.

\bibitem[\protect\citeauthoryear{Vidakovic}{Vidakovic}{2009}]{vidakovic:2009}
Vidakovic, B. (2009).
\newblock {\em Statistical Modeling by Wavelets}.
\newblock Hoboken: John Wiley and Sons.

\bibitem[\protect\citeauthoryear{Wand and Jones}{Wand and Jones}{1994}]{wand:1994}
Wand, M.~P. and M.~C. Jones (1994).
\newblock {\em Kernel smoothing}.
\newblock New York: CRC press.

\bibitem[\protect\citeauthoryear{White, Frye, Christensen, Gelfand, and Silander}{White et~al.}{2022}]{white:2022}
White, P.~A., H.~Frye, M.~F. Christensen, A.~E. Gelfand, and J.~A. Silander (2022).
\newblock Spatial functional data modeling of plant reflectances.
\newblock {\em The Annals of Applied Statistics\/}~{\em 16\/}(3), 1919--1936.

\bibitem[\protect\citeauthoryear{White, Keeler, and Rupper}{White et~al.}{2021}]{Philip:2021}
White, P.~A., D.~G. Keeler, and S.~Rupper (2021).
\newblock {Hierarchical Integrated Spatial Process Modeling of Monotone West Antarctic Snow Density Curves}.
\newblock {\em The Annals of Applied Statistics\/}~{\em 15\/}(2), 556--571.

\bibitem[\protect\citeauthoryear{Zhang, Baladandayuthapani, Zhu, Baggerly, Majewski, Czerniak, and Morris}{Zhang et~al.}{2016}]{zhang:2016}
Zhang, L., V.~Baladandayuthapani, H.~Zhu, K.~A. Baggerly, T.~Majewski, B.~A. Czerniak, and J.~S. Morris (2016).
\newblock Functional {CAR} models for large spatially correlated functional datasets.
\newblock {\em Journal of the American Statistical Association\/}~{\em 111\/}(514), 772--786.

\bibitem[\protect\citeauthoryear{Zhou, Huang, Martinez, Maity, Baladandayuthapani, and Carroll}{Zhou et~al.}{2010}]{Zhou:2010}
Zhou, L., J.~Z. Huang, J.~G. Martinez, A.~Maity, V.~Baladandayuthapani, and R.~J. Carroll (2010).
\newblock {Reduced Rank Mixed Effects Models for Spatially Correlated Hierarchical Functional Data}.
\newblock {\em Journal of the American Statistical Association\/}~{\em 105\/}(489), 390--400.

\end{thebibliography}

\appendix
\section{Measures of Discrepancy} \label{sec:app_metrics}

An estimator for a function evaluated at a specific point shares similarities with an estimator for a scalar parameter. Key characteristics of these estimators include their expectations and various properties related to random variables. It's crucial to understand that when discussing an expectation (denoted by $\mathbb{E}[\cdot]$) or variance ($\mathbb{V}[\cdot]$), these metrics are computed based on the (unknown) distribution of the associated random variable. We often utilize the empirical distribution to derive these values in practical applications. To effectively evaluate the performance of a function estimator, it is vital to establish appropriate criteria for measuring estimation errors. When focusing on the estimation at a single point $t$, a commonly used metric is the Mean Square Error (MSE), defined as $\text{MSE}[\widehat{Y}(t)] = \mathbb{E}[\widehat{Y}(t)-Y(t)]^2$. We can further break down and analyze the MSE by applying basic principles of mean and variance.
\begin{align} \label{ec:2.16} 
 \text{MSE}[\widehat{Y}(t)] \ = \ [\mathbb{E}[\widehat{Y}(t)]-Y(t)]^{2} \ + \ \mathbb{V}[\widehat{Y}(t)]. 
\end{align}

Instead of simply estimating $Y(t)$ at a fixed point, evaluating the function over the entire real line is often desirable, especially from a data analysis viewpoint. In this case, the estimate is the function $\widehat{Y}(t)$, so it is necessary to consider an error criterion that globally measures the distance between the functions $\widehat{Y}(t)$ and $Y(t)$. Generally, these criteria can be defined as a norm of the function. The $L^{p}$ norm \citep{gentle:2009, wand:1994} of the error is $(\int_{T}| \widehat{Y}(t) -Y(t)|^{p} dt)^{1/p}$, where $T$ is the domain of $Y$ (true function). The estimator $\widehat{Y}(t)$ must also be defined over the same domain. The integral may not exist. Two useful measures are the $L^{2}$ norm, also called the integrated squared error (ISE) 
\begin{align}
    \text{ISE} (\widehat{Y}(t))=\int_{T} (\widehat{Y}(t) -Y(t))^{2} dt.
\end{align}

\section{Missing Data}\label{app:B}
\medskip
This simulation study evaluates the performance of the proposed model in handling missing data in time series corresponding to different locations. Unlike the analysis presented in Subsection \ref{sec:3.2}, in which the series of a specific location $s_{j}$ was omitted entirely, in this study, we assume that only some observations are missing in each series. To do so, we start from the data in Subsection \ref{sec:3.2} and randomly remove some observations, thus generating samples with missing data. Notably, the location of the missing data is kept constant in all MC replications. This design includes 750 missing values distributed among the 15 curves, representing 25$\%$ of the 3,000 observations in each MC sample.

\begin{figure}[!hbt]	
	\centering
	\includegraphics[width=1\textwidth]{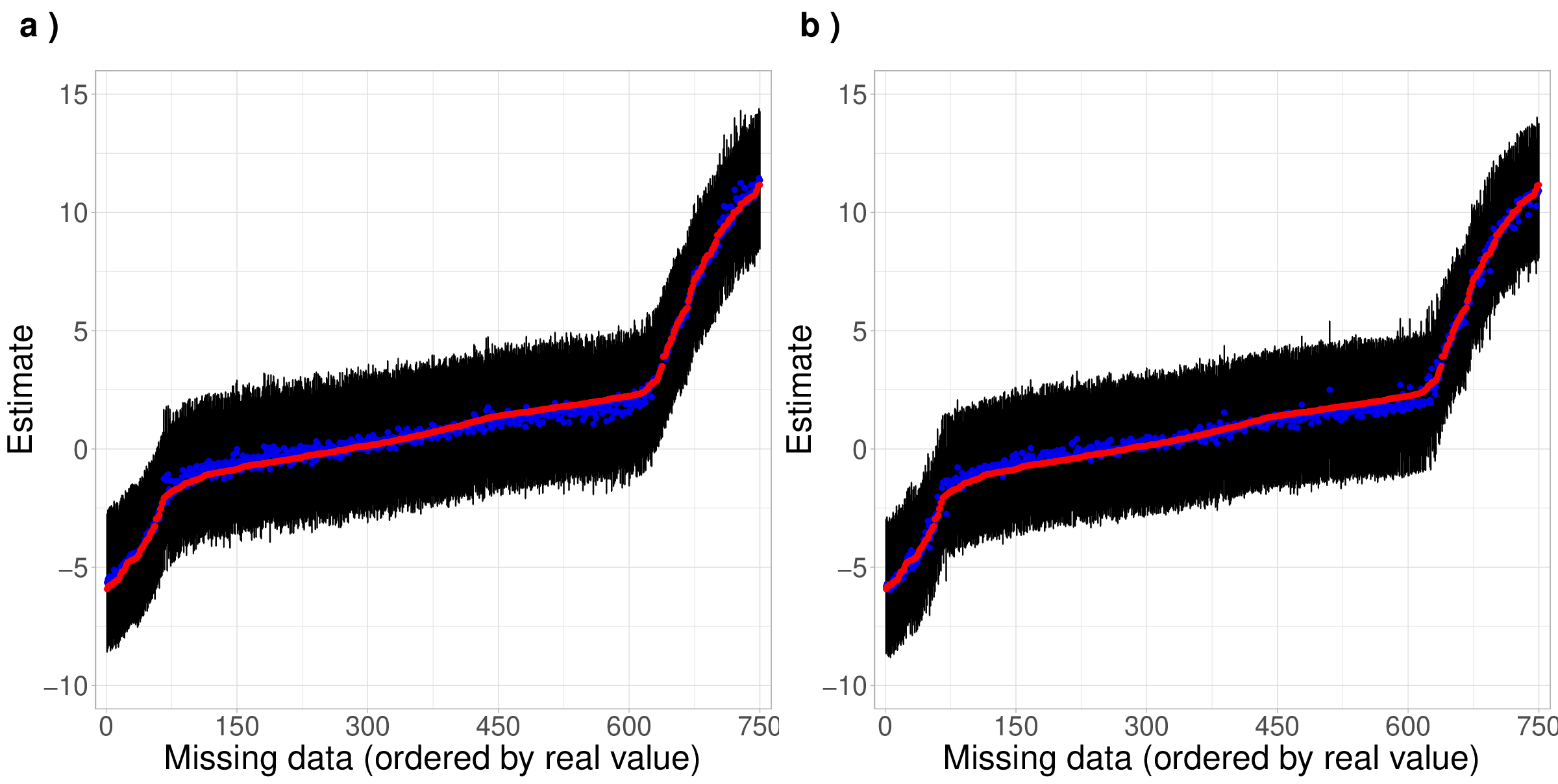}
	\caption{Average (100 replicas) 95$\%$ HPD Intervals for the 750 missing data using the $\mathcal{M}_{BP_{17};\,\delta}$. The real values are the red dots. The average (100 replicas) posterior means are the blue dots: Panel (a) 12 functional basis. Panel (b) 18 basis.}
	\label{fig:B}
\end{figure}

Figure \ref{fig:B} illustrates the 95\% HPD intervals for the 750 missing values in the MC datasets. The blue dots represent the mean posterior estimates, while the red dots indicate the target observations. The graphs are arranged according to the actual values to enhance visual analysis. It is important to note that these graphs do not represent curves in the functional domain. The study examines two scenarios with different numbers of basis functions: 12 and 18. Both panels show that the HPD intervals' MC means effectively capture the missing data's target values. Although the intervals exhibit a degree of uncertainty, most mean posterior estimates from each MC replicate align closely with the target values, primarily when the proposed model utilizes 18 basis functions. In this case, the intervals are narrower than those in Panel (a), indicating greater precision in the estimates.
\newpage
\end{document}